\documentclass[prc,showpacs,showkeys,superscriptaddress,nofootinbib,floatfix,twocolumn]{revtex4}
\usepackage{color}
\usepackage{amsfonts}
\usepackage{amsbsy}
\usepackage{mathrsfs}
\usepackage{graphicx}
\usepackage{subfigure}
\newcommand{\eeq}{\end{equation}}
\newcommand{\bea}{\begin{eqnarray}}
\newcommand{\eea}{\end{eqnarray}}

\def\lsim{\mathrel{\rlap{
\lower4pt\hbox{\hskip-3pt$\sim$}}
    \raise1pt\hbox{$<$}}}     
\def\gsim{\mathrel{\rlap{
\lower4pt\hbox{\hskip-3pt$\sim$}}
    \raise1pt\hbox{$>$}}}     

\begin{document}





\title{Color path-integral Monte-Carlo simulations \\ 
of quark-gluon plasma: Thermodynamic and transport properties
  }%
\author{V.S.~Filinov}
\thanks{Corresponding author\quad E-mail:~\textsf{vladimir\_filinov@mail.ru}}
\affiliation{Joint Institute for High Temperatures, Russian Academy of
Sciences, Moscow, Russia}
\author{Yu.B. Ivanov}
\affiliation{Kurchatov Institute,
Kurchatov sq. 1,
Moscow, Russia}
\author{V.E. Fortov}
\affiliation{Joint Institute for High Temperatures, Russian Academy of
Sciences, Moscow, Russia}
\author{M. Bonitz}
\affiliation{Institute for Theoretical Physics and Astrophysics, Christian Albrechts University, 
Kiel, Germany}
\author{P.R. Levashov}
\affiliation{Joint Institute for High Temperatures, Russian Academy of
Sciences, Moscow, Russia}
%


\begin{abstract}

Based on the quasiparticle model of the quark-gluon plasma (QGP), a
color quantum path-integral Monte-Carlo (PIMC) method for calculation of thermodynamic properties
and -- closely related to the latter -- a Wigner dynamics method for calculation of transport properties
of the QGP are formulated.
The QGP partition function is
presented in the form of a color path integral with a new relativistic measure instead of the Gaussian one traditionally used in the
Feynman-Wiener path integral.  A procedure of sampling color variables according to the SU(3) group Haar measure is developed for integration
over the color variable. It is shown that the PIMC method
is able to reproduce the lattice QCD equation of state at zero baryon chemical potential at
realistic model parameters (i.e. quasiparticle masses and coupling constant)
and also yields valuable insight into the internal structure of the QGP.
Our results indicate that the QGP reveals {\em quantum liquid-like}
(rather than gas-like) properties up to the highest considered temperature of 525 MeV.
The pair distribution functions clearly reflect the existence of gluon-gluon bound states, i.e. glueballs,
at temperatures just above the phase transition, while
meson-like $q\overline{q}$ bound states are not found.
The calculated self-diffusion coefficient agrees well with some estimates of the heavy-quark diffusion constant
available from recent  lattice
data and also with an analysis of heavy-quark quenching in experiments
on ultrarelativistic heavy ion collisions, however,
appreciably exceeds other estimates.
The lattice and heavy-quark-quenching results
on the heavy-quark diffusion  are still rather diverse.
The obtained results for the shear viscosity are in the range of those
deduced from an analysis of the experimental elliptic flow in ultrarelativistic heavy ions collisions,
i.e.  in terms the viscosity-to-entropy ratio,
$1/4\pi\lsim \eta/S < 2.5/4\pi$, in the temperature range from 170 to 440 MeV.
\end{abstract}


\maketitle

\section{Introduction}\label{s:intro}

Determining the properties of the quark-gluon plasma
(QGP) is nowadays one of the most important goals in
high-energy nuclear physics.
In recent years, experiments at the Relativistic Heavy-Ion Collider
(RHIC) at Brookhaven National Laboratory \cite{shuryak08}
and the Large Hadron Collider (LHC) at CERN \cite{Schukraft:2011ch}
have provided a wealth of data from which one can obtain
information on a number of features of the QGP.
The most striking result, obtained from analysis
of these experimental data \cite{shuryak08,Shen:2012vn}, is that the deconfined
quark-gluon matter behaves as an almost perfect fluid rather than a perfect gas,
as it could be expected from the asymptotic freedom.

There are  various approaches for a theoretical study of the QGP each of which has its advantages and  disadvantages.
The most fundamental way to compute
the properties of strongly interacting matter is provided by the lattice QCD \cite{
Fodor09,Csikor:2004ik,Cheng:2009zi}.
Interpretation of these very complicated computations
requires application of various QCD motivated, albeit schematic, models simulating various aspects of the full theory.
Moreover, such models are needed in cases when the lattice QCD fails, e.g. at large
quark chemical potentials and out of equilibrium.
While certain progress has been achieved in recent years, transport properties
of the QGP are still poorly accessible within lattice QCD -- only viscosity for pure
gluonic matter was computed \cite{Meyer:2007ic}.
It is, therefore, crucial to devise reliable and manageable theoretical tools for a quantitative description of non-Abelian QGP both in and out of equilibrium.

A semi-classical approximation, based on a quasiparticle picture has been introduced in Refs. \cite{Levai,LM105,LM109,LM110,LM111}.
It is motivated by the expectation that the main features of non-Abelian
plasmas can be understood in simple semi-classical terms without the
 difficulties inherent to a full quantum field-theoretical analysis.
Independently the same ideas were implemented in terms of molecular dynamics (MD) \cite{Bleicher99};
this approach was further developed in a series of works
\cite{shuryak1,Zahed}. The MD allows one to treat soft processes in the QGP which
are not accessible by perturbative means.

Strongly correlated behavior of the QGP is expected to show up in long-ranged spatial correlations of quarks and
gluons which, in fact, may give rise to liquid-like and, possibly, solid-like structures.
This expectation is based on a very similar
behavior observed in electrodynamic plasmas  \cite{shuryak1,thoma04,afilinov_jpa03}. This similarity was exploited
to formulate a classical nonrelativistic model of a color Coulomb interacting QGP \cite{shuryak1} which was  numerically analyzed by classical MD simulations.
Quantum effects were either neglected or incorporated
phenomenologically via a short-range repulsive correction to the pair potential. Such a rough model may, however, become
a critical issue at high densities.
Similar models in electrodynamic plasmas showed poor behavior in the region of strong wave function overlap, in particular around the Mott density where bound states break up.
For temperatures and densities of the QGP considered
in  Ref. \cite{shuryak1} these effects are very important since the quasiparticle
thermal wave length is of the order of the average interparticle distance.
Therefore, to account for quantum effects, we follow an idea of Kelbg \cite{kelbg} that allows one to  rigorously include quantum corrections to the pair potential\footnote{
The idea to use a Kelbg-type effective potential also for quark matter
was independently proposed  by K.~Dusling
and C.~Young \cite{dusling09}. However, their potentials are limited to weakly nonideal systems.
}.
Strictly speaking, this method is applicable only to weak coupling.
To extend the method  to stronger couplings, an ``improved Kelbg potential''
was derived, which contains a single free parameter,
being fitted to the exact solution of the quantum-mechanical two-body problem.
Using the method of the improved Kelbg potential in classical MD simulations one is able to describe
thermodynamic properties of a partially ionized plasma up to moderate couplings \cite{afilinov_pre04}.
However, this approach may fail, if bound states of more than two
particles are formed in the system. This is a result of
break-down of the pair approximation for the density matrix, as demonstrated in Refs. \cite{afilinov_pre04}. A superior approach, which does not have this limitation, consists in the use of the original Kelbg potential in path integral Monte Carlo
simulations (PIMC) which effectively map the problem onto a high-temperature weakly coupled and weakly degenerate one.
This allows one to advance the analysis to strong couplings and is, therefore, a relevant choice for the present purpose.
The PIMC method has been successfully applied to various phases of strongly coupled electrodynamic plasmas
\cite{filinov_ppcf01,bonitz_jpa03}. 
Examples are partially-ionized dense hydrogen plasmas, where liquid-like and
crystalline behavior was observed \cite{filinov_jetpl00,bonitz_prl05,
filinov_jpa03}, as well as electron-hole plasmas in
semiconductors \cite{bonitz_jpa06,filinov_pre07}, including excitonic bound states.

In this paper we extend the previous classical nonrelativistic simulations \cite{shuryak1} in two ways: first, we
include quantum and spin effects and, second, we take into account the dominant relativistic effects, cf. section \ref{Thermodynamics}.
This is done in the frame of quantum Monte Carlo simulations where
we rewrite the partition function of this system in the form of color path integrals with a new relativistic measure instead of Gaussian one used in Feynman-Wiener path integrals. For the integration of the partition function over color variables we develop a procedure of sampling the color quasiparticle variables according to the SU(3) group Haar measure with the quadratic and cubic Casimir conditions. The developed approach self-consistently takes  into account the Fermi (Bose) statistics of quarks (gluons).
The main goal of this article is to test the present Color Path-Integral Monte-Carlo (PIMC)
against known lattice data \cite{Fodor09} and to predict additional properties of the QGP, which are still
inaccesible from lattice QCD.
First results of  the path integral approach for thermodynamic properties of the nonideal QGP
have been already briefly reported in Ref. \cite{Filinov:2012pt}
for SU(3) group and
in Refs. \cite{Filinov:2012pimc,Filinov:2010pimc,Filinov:2012zz,Filinov:2009pimc} for SU(2) group.
In this paper we  show that the PIMC method is
able to reproduce the QCD lattice equation of state at vanishing baryon-charge density
and also yields valuable insight into the internal structure of the QGP.
These results are presented in section \ref{e:model}.

Hydrodynamic simulations of relativistic heavy-ion collisions require not only knowledge of thermodynamic
 properties of the QGP but also of the transport properties.
Unfortunately the PIMC method itself is not able to directly predict transport properties.
Therefore, to simulate quantum QGP transport and thermodynamic properties
within a unified approach we combine the path integral and Wigner (in phase space) formulations
of quantum mechanics (section \ref{Wigner}).
There the  kinetic  coefficients are calculated by means of Kubo formulas. In this approach the PIMC method is used
to generate initial conditions  (equilibrium quasiparticle configurations) for
dynamical trajectories describing the time evolution for spatial, momentum and color variables.
Correlation functions and kinetic coefficients are calculated as averages
 of Weyl's symbols of dynamic operators along these trajectories.
The basic ideas of this approach have been published in Ref. \cite{ColWig11}.
{ 
This method is applicable to systems with arbitrary
strong interaction.}
Using this approach we calculate the self-diffusion coefficient and  viscosity of the strongly coupled QGP.
These results are presented in section \ref{k:model}.

\section{Thermodynamics of QGP} \label{Thermodynamics}
In this section we summarize the main ideas of our approach to  the thermodynamic properties of the strongly correlated quark-gluon plasma.
This approach is based on a generalization of the Feynman path integral representation of quantum mechanics to high energy matter. Before
deriving the main equations of our PIMC approach we specify the simplifications and model parameters.

\subsection{Basics of the model}\label{semi:model}

The basic assumptions of the model are similar to those of  Ref. \cite{shuryak1}:
\begin{description}
 \item[I.]
 Quasiparticles masses ($m$) are of order or higher
 than the mean kinetic energy  per particle. This assumption is based on the analysis of QCD lattice
 data \cite{Lattice02,LiaoShuryak,Karsch:2009tp}. For instance, at zero net-baryon density it
 amounts to $m \gsim T$, where $T$ is a temperature.
 \item[II.] 
 In view of the first assumption,
 interparticle interaction is dominated
 by a color-electric Coulomb potential.
 Magnetic effects are neglected as subleading ones. 
  \item[III.] Relying on the fact that the color representations are large, the color operators 
 are substituted by their average values, i.e. by Wong's classical color vectors (8D in SU(3))
 with the quadratic and cubic Casimir conditions \cite{Wong}.
 \item[IV.] We consider the 3-flavor quark model.
  {  
  For the sake of simplicity we assume the masses
  of '$u$p', '$d$own' and '$s$trange' quarks
  to be equal.}
 As for the gluon quasiparticles, we allow their mass to be different (heavier) from that of quarks.
\end{description}
The quality of these approximations and their limitations were discussed in  Ref. \cite{shuryak1}.
Thus, 
this model requires the following quantities as a function of temperature ($T$)
and quark chemical potential ($\mu_q$) as an input:
\begin{description}
\item[1.] quasiparticle masses, for quarks $m_q$ and gluons $m_g$, and
\item[2.] the coupling constant $g^2$, or $\alpha_s = g^2/4\pi$.
\end{description}
Input quantities should be deduced from lattice QCD data or from an appropriate model simulating these data.

It has been established that hard modes (in terms of {\em hard thermal loop} approximation
\cite{Pisarski89,Braaten90,Blaizot02}) behave like quasi-particles \cite{Blaizot02}.
Therefore, masses of these quasiparticles should be deduced from nonperturbative calculations
taking into account hard field modes,
e.g., they can be associated with pole masses deduced
from lattice QCD calculations.
At the same time, the soft quantum
fields are characterized by very high occupation numbers per mode. Therefore,
to leading order, they can be well approximated by soft classical fields.
This is precisely the picture we are going to utilize: massive
quantum quasiparticles (hard modes) interacting via classical color fields.
Applicability of such approach was discussed in Refs. \cite{LM105,shuryak1} in detail. Our approach
differs from that of Ref. \cite{LM105,shuryak1} by a quantum treatment to quasiparticles instead
of the classical one, and additionally by a relativistic description of the kinetic energy instead of the nonrelativistic
approximation of Ref. \cite{shuryak1}.

\subsection{Color Path Integrals}\label{s:pimc}
We consider a multi-component QGP consisting of $N$ color quasiparticles:  $N_g$  gluons,  $N_q$  quarks and $\overline{N}_{q}$ antiquarks. The Hamiltonian of this system is
 $\hat{H}=\hat{K}+\hat{U}^C$ with
 the kinetic and color Coulomb interaction parts
\begin{eqnarray}
\label{Coulomb}
{\hat{K}}&=&\sum_i \sqrt{\hat{\bf p}^2_i+m^2_i(T,\mu_q)},
\cr
{\hat{U}^C}&=&\frac{1}{2}\sum_{i\neq j}
\frac{g^2(T,\mu_q)
(Q_i\cdot Q_j)}
{4\pi| {\bf r}_i- {\bf r}_j|}.
\end{eqnarray}
Here $i$ and $j$ summations run over quark and gluon quasiparticles, $i,j=1,\ldots,N$,
$N=N_q+\overline{N}_q+N_g$,
$N_q=N_u+N_d+N_s$ and $\overline{N}_{q}=\overline{N}_{u}+\overline{N}_{d}+\overline{N}_{s}$ are total
numbers of quarks and antiquarks of all  flavours ({\em u}p, {\em d}own and {\em s}trange), respectively,
3D vectors ${\bf r}_i$ are quasiparticle spatial coordinates, the $Q_i$ denote
the Wong's quasiparticle color variable (8D-vector in the group $SU(3)$),
$(Q_i\cdot Q_j)$ denote scalar product of color vectors.
Nonrelativistic approximation for potential energy is used,
while for kinetic energy we still keep relativistic form as the quasiparticle masses are not negligible
as compared with temperature. The eigenvalue equation of this Hamiltonian is usually called
the ⌠spinless Salpeter equation.■ It may be regarded as a well-defined approximation to the Bethe√Salpeter formalism
 \cite{SB1} for the description of bound states within relativistic quantum field theories, obtained when assuming
that all bound-state constituents interact instantaneously and propagate like free particles \cite{SB2}.
Among others, it yields semirelativistic descriptions of hadrons as bound states of quarks
\cite{LS1,LS2}.

In the classic approximation this system is governed by Wong's equations of motion \cite{Wong}
\bea 
\frac{d{\bf p}_i(t)}{dt}&=& {\bf F}_i(t) , 
\label{Wong-eom-p}
\\
\frac{d{\bf r}_i(t)}{dt}&=& {\bf v}_i(t),  
\label{Wong-eom-r}
\\
\frac{dQ^a_i(t)}{dt}&=& \Upsilon^a_i(t), 
\label{Wong-eom-Q}
\eea
where ${\bf p}_i$ is the momentum of a quasiparticle,
${\bf v}_i={\bf p}_i/\sqrt{{\bf p}_i^2+m_i^2}$ is its velocity,
${\bf F}_i = -\partial U^C/\partial {\bf r}_i$ is the  color-electric force experienced by the quasiparticle,
and
\bea
\Upsilon^a_i=\sum^N_{j\neq i} \sum_{b,c} \breve{f}_{abc}
\frac{g^2 Q^b_i Q^c_j } {4\pi|{\bf r}_i-{\bf r}_j|},
\label{R-Wong}
\eea
is the driving force in equation of motion for the color charge,
$\breve{f}_{abc}$ are structure constants of the group SU(3) and $a,b,c =1,\dots,8$ (see Appendix I).
Below we consider spatial degrees of freedom quantum-mechanically while the color dynamics, still classically.

Thermodynamic properties in the grand canonical ensemble with given temperature ($T$),
net-quark-number ($\mu_q$) and strange ($\mu_s$) chemical potentials,
and fixed volume $V$ are fully described by the
grand partition function

%
\begin{widetext}
\begin{eqnarray}
Z\left(\mu_q,\mu_s,\beta,V\right)
&=&
\sum_{\{N\}}
\frac{\exp\{\mu_q(N_q-\overline{N}_{q})/T\}\;\exp\{\mu_s(N_s-\overline{N}_{s})/T\}}%
{N_u!\;N_d!\;N_s!\;\overline{N}_{u}!\;\overline{N}_{d}!\;\overline{N}_{s}!\;N_g!}
Z\left(\{N\},V,\beta\right) \text{,}
\label{Gq-def}
\\
Z\left(\{N\},V,\beta\right) &=& \sum_{\sigma} \int\limits_V
dr\; d\mu Q \;\rho(r,Q, \sigma; \{N\}; \beta),
\label{Z-def}
\end{eqnarray}
\end{widetext}
where $\{N\}=\{N_u,N_d,N_s,\overline{N}_{u},\overline{N}_{d},\overline{N}_{s},N_g\}$,
$\rho(r,Q, \sigma;\{N\} ; \beta)$ denotes the diagonal matrix
elements of the density-matrix operator
${\hat \rho} = \exp (- \beta{\hat H})$ with $\beta=1/T$.
Here $r$, $\sigma$ and  $Q$  denote the multi-dimensional vectors related  spatial, spin and color
degrees of freedom, respectively, of all quarks, antiquarks and gluons.
The $\sigma$ summation and  spacial ($dr\equiv d^3 r_1 ...d^3 r_N $)
and color ($d\mu Q\equiv d\mu Q_1 ... d\mu Q_N $) integrations
run over all individual degrees of freedom of the particles,
$d\mu Q_i$ denotes integration over SU(3) group Haar measure, see Appendix I.
Usual choice of the strange  chemical potential is $\mu_s=-\mu_q$ (nonstrange matter),
such that the total factor in front of $(N_s-\overline{N}_{s})$ is zero.
Therefore, below we omit $\mu_s$ from the list of variables. 
In Eq. (\ref{Gq-def}) we explicitly wrote sum over different quark flavors (u,d,s).
Below the sum over quark degrees of freedom is understood in the same way.

Since the masses and the coupling constant depend on the temperature and quark chemical potential,
special care should be taken to preserve thermodynamical consistency of this approach.
To achieve this,
thermodynamic functions such as pressure, $P$, entropy, $S$, baryon number, $N_B$, and
internal energy, $E$, should be calculated through respective derivatives of
 the logarithm of the partition function
\begin{eqnarray}
\label{p_gen}
P&=&\partial (T\ln Z) / \partial V, 
\\
\label{s_gen}
S&=&\partial (T\ln Z) / \partial T, 
\\
\label{n_gen}
N_B&=&(1/3)\partial (T\ln Z) / \partial \mu_q, 
\\
\label{e_gen}
E&=& -PV+TS+3 \mu_q N_B.
\end{eqnarray}
This is a conventional way of maintaining the thermodynamical consistency in approaches
of the Ginzburg--Landau type as they are applied in high-energy physics, e.g., in the PNJL model.

The exact density matrix $\rho=e^{-\beta {\hat H}}$ of interacting quantum
systems can be constructed using a path integral
approach~\cite{feynm,zamalin}
based on the operator identity
\begin{eqnarray}
\label{Texp}
e^{-\beta \hat{H}}= e^{-\Delta \beta {\hat H}} \cdot
e^{-\Delta \beta {\hat H}}\dots  e^{-\Delta \beta {\hat H}}
\end{eqnarray}
where the r.h.s. contains $n+1$ identical factors with $\Delta \beta = \beta/(n+1)$,
which allows us to
rewrite
the integral in Eq.~(\ref{Z-def}) in the form
\begin{widetext}
\begin{eqnarray}
&&
\sum_{\sigma } \int\limits dr^{(0)}d\mu Q\,
\rho(r^{(0)},Q,\sigma ; \{N\};\beta)
\nonumber\\&=&
\sum_{\sigma} \int\limits  d\mu Q \, dr^{(0)} dr^{(1)}\dots
dr^{(n+1)} \, \rho^{(1)} \rho^{(2)} \, \dots \rho^{(n)}
\nonumber\\&\times&
\sum_{P_q} \sum_{P_{ \overline{q}}}\sum_{P_g}(- 1)^{\kappa_{P_q}+ \kappa_{P_{\overline{q}}}}
{\cal S}(\sigma^{(0)}, {\hat P_q}{\hat P_{ \overline{q}}}{\hat P_g} \sigma^{(n+1)})\,
{\hat P_q} {\hat P_{ \overline{q}}}{\hat P_g}\rho^{(n+1)}\big|_{\sigma^{(n+1)}=\sigma^{(0)}}
\delta\left(r^{(n+1)}- r^{(0)}\right)
\nonumber\\
&\equiv&
\int\limits d\mu Q\int dr^{(0)} dr^{(1)}\dots dr^{(n+1)}
{R}\left(r^{(0)},r^{(1)}, \dots ,r^{(n+1)};Q;\{N\};\beta\right)\;\delta\left(r^{(n+1)}- r^{(0)}\right).
 \label{Grho-pimc}
\end{eqnarray}
\end{widetext}
For the sake of notation convenience, here we ascribe superscript $^{(0)}$
to the original variables.
Notice that the color charge $Q$ is a classical variable already in the mixed
(i.e. coordinate-momentum) representation, see Appendix I. Therefore, we do not
build a chain of $n$ different $Q$-variables.
The spin variable $\sigma$ is the same in all $\rho^{(l)}$ except for the
 $\rho^{(n+1)}$,where it  is initially set $\sigma^{(n+1)}$ and only after
  permutations performed is put to $\sigma$.
The spin gives rise to the spin part of the density matrix (${\cal
S}$).
{ 
To take into account the Fermi/Bose statistics of (anti)quarks/gluons it is necessary
to antisymmetrize/symmetrize the density matrix over respective spatial, color, 
and spin variables.
In the product of $\rho^{(l)}$ it is enough to perform this
antisymmetrization/symmetrization only in a single term  \cite{huang},
since in fact the variables of any $\rho^{(l)}$ are related to the same
set of quasiparticles.
We choose it to be $\rho^{(n+1)}$. The (anti)symmetrization is done by
the permutation operators  $\hat P_q$, $\hat P_{ \overline{q}}$ and
$\hat P_g$ acting on related 
spatial $r^{(n+1)}$,   
spin $\sigma^{(n+1)}$ and color  $Q$ variables  in  $\rho^{(n+1)}$.
The sum runs over all permutations with parity factors $\kappa_{P_q}$ and
$\kappa_{P_{ \overline{q}}}$ corresponding to each permutation.}
In Eq.~(\ref{Grho-pimc})
\begin{eqnarray}
\rho^{(l)}
&=&
\rho\left(r^{(l-1)},r^{(l)},Q;\{N\};\Delta\beta\right) 
\cr
&=&
\left\langle r^{(l-1)},Q\left|e^{-\Delta \beta {\hat H}}\right|r^{(l)},Q\right\rangle,
\label{rho(l)}
\end{eqnarray}
is an off-diagonal element of the density matrix.
Accordingly each quasiparticle is represented by a set of coordinates
$\{r_i^{(0)}, \dots , r_i^{(n)}\}$ (``beads'')
and a 8-dimensional color
vector $Q_i$ in the $SU(3)$ grup.
Thus, all "beads" of each quasiparticle are characterized by the same spin projection,  flavor and color charge. Notice that masses and coupling
constant in each $\rho^{(l)}$ are the same as those for the original quasiparticles, i.e. these are still defined by the actual temperature $T$.

The main advantage of decomposition (\ref{Grho-pimc}) is that it
allows us to use perturbation theory to obtain approximation for density matrices $\rho^{(l)}$,
which is applicable due to smallness of artificially introduced factor $1/(n+1)$.
This means that
in each $\rho^{(l)}$  the ratio
$g^2(T,\mu_q) (Q_i\cdot Q_j) /[4\pi| r^{(l)}_i- r^{(l)}_j|T(n+1)]$
can be always made much smaller than one,
which allows us to use perturbation theory with respect to the potential.
Each factor in Eq. (\ref{Grho-pimc}) should be calculated with the accuracy of order of $1/(n+1)^\theta$
with $\theta > 1 $, as in this case the error of the whole product in the limit of large $n$
will be equal to zero.
In the limit $(n+1)\longrightarrow \infty$, $\rho^{l}$ can be approximated by a product of two-particle density matrices $\rho_{ij}^{(l)}$ \cite{filinov_ppcf01,feynm,zamalin}.
{
This approximation can be deduces from operator expansion
\begin{eqnarray}
&&\exp \left(- \Delta\beta{\hat H}\right)
\cr
&\approx&
\exp \left(- \frac{\Delta\beta}{2}{\hat K}\right)
\exp \left(- \Delta\beta{\hat U^C}\right)
\exp \left(- \frac{\Delta\beta}{2}{\hat K}\right)
\cr
&\times&
\left\{\mbox{terms with  } (\frac{\Delta\beta}{2})^2[{\hat K},{\hat U^C}]\ldots \right\}
\, . \label{Morita}
\end{eqnarray}
As the first approximation with the error proportional to $1/(n+1)^2$ ,  we can write
\begin{eqnarray}
&&\exp \left(- \Delta\beta{\hat H}\right)
\cr
&\approx&
\exp \left(- \frac{\Delta\beta}{2}{\hat K}\right)
\exp \left(- \Delta\beta{\hat U}^C\right)
\exp \left(- \frac{\Delta\beta}{2}{\hat K}\right)
\, , \label{Mori}
\end{eqnarray}
or a simpler expression based on neglecting the contribution of commutator $\left[{\hat K},{\hat U}^C\right]$
\begin{eqnarray}
\exp \left(- \Delta\beta{\hat H}\right)\approx
\exp \left(- \Delta\beta{\hat K}\right)
\exp \left(- \Delta\beta{\hat U}^C\right)
\, . \label{Morit}
\end{eqnarray}
In order to save computational time and resources we use
the simpler expression (\ref{Morit}) for calculation of the thermodynamic
quantities. This approximation has the same order of the error as that in Eq. (\ref{Mori})
but contain a larger numerical coefficient in front of $1/(n+1)^2$.
For calculations of transport properties we use approximation (\ref{Mori}) as it will be explained in Section \ref{s:InCd}.}

{
It is very important that in both approximations the error of the whole product in Eq. (\ref{Grho-pimc}) is
proportional to $1/(n+1)$ and tends to zero in the limit of $n\rightarrow\infty$.
The second advantage of the decomposition of Eq. (\ref{Grho-pimc}) is that it reduces quantum
multi-particle interaction to the pair-wise sum of two-particle  interactions described by
two-particle classical density matrices  in each $\rho^{(l)}$.
}

Thus, neglecting the commutator terms in Eq. (\ref{Morita}), we arrive at the following expression for the density matrix of Eq. (\ref{rho(l)})
\begin{eqnarray}
\rho^{(l)}_{\rm neglecting \; commutators}
=
\rho^{(l)}_{0}\, \exp\left[-\Delta\beta U^C\left(r^{(l)},Q\right)\right]
\label{negl.comm.}
\end{eqnarray}
where $\rho^{(l)}_{0}$ is the corresponding density matrix of noninteracting particles.
This approximation works well for potentials bounded below. However, the Coulomb potential
can go to minus infinity and hence the result (\ref{negl.comm.}) diverges in this limit.

A more sophisticated treatment is required to avoid this divergence.
All the calculations along this line can be rigorously performed for the two-particle density matrix
$\rho^{[2]} (r,r',Q;\Delta\beta)$, where
$r = \{{\bf r}_1, {\bf r}_2\}$, $r' = \{{\bf r}'_1, {\bf r}'_2\}$ and $Q = \{Q_1, Q_2\}$.
Expanding the two-particle density matrix
up to the second order in $1/(n+1)$, one arrives to the following result  \cite{kelbg}
\begin{eqnarray}
&&
\rho^{[2]}(r,r',Q;\Delta\beta)
\approx
\rho^{[2]}_0(r,r',Q;\Delta\beta)
\nonumber\\&-&
\int_0^1 d\tau \int d^3 \widetilde{r}
\frac{\Delta\beta \, g^2(Q_i\cdot Q_j)}{4\pi|\widetilde{\bf r}|\Delta\lambda_{12}^2\sqrt{\tau(1-\tau)}}
\nonumber\\&\times&
\exp\left(-\frac{\pi|{\bf r}_{12}-\widetilde{\bf r}|^2}{\Delta\lambda_{12}^2(1-\tau)}\right)
\exp\left(-\frac{\pi|\widetilde{\bf r}-{\bf r'}_{12}|^2}{\Delta\lambda_{12}^2\tau}\right)
\nonumber\\&\approx&
 \rho^{[2]}_0
\exp\left[-\Delta\beta \,
\Phi_{12}({\bf r}_{12},{\bf r'}_{12},Q_1,Q_2)\right],
\label{GPERT}
\end{eqnarray}
where ${\bf r}_{12}={\bf r}_{1}-{\bf r}_{2}$, ${\bf r'}_{12}={\bf r'}_{1}-{\bf r'}_{2}$,
$\Delta\lambda_{12}=\sqrt{2\pi\Delta\beta /m_{12}}$ is defined in terms of
the reduced mass of the pair of particles: $m_{12}=m_{1}m_{2}/(m_{1}+m_{2})$,
and $\rho^{[2]}_0$ is
the two-particle density matrix of noninteracting particles.
In the end of Eq. (\ref{GPERT}) the result is presented in the form similar to Eq. (\ref{negl.comm.}),
i.e. in terms of an off-diagonal two-particle effective quantum potential $\Phi_{12}$,
which is called a Kelbg potential \cite{kelbg}. Eq. (\ref{GPERT}) is the definition of the color Kelbg potential.
The diagonal part of the color Kelbg potential can be obtained analytically
\begin{eqnarray}
&&\Phi_{12}({\bf r}_{12},{\bf r}_{12},Q_1,Q_2) 
\approx \frac{g^2\,(Q_i\cdot Q_j)}{4 \pi \Delta\lambda_{12} x_{12}} 
\cr
&\times&
\left\{1-e^{-(x_{12})^2} +
\sqrt{\pi} x_{12} \left[1-{\rm erf}(x_{12})\right] \right\},
\label{kelbg-d}
\end{eqnarray}
where $x_{12}=|{\bf r}_{12}|/\Delta\lambda_{12}$.
Notice that the color Kelbg
potential approaches the color  Coulomb potential
at distances larger than $\Delta\lambda_{12}$. What is of prime importance, the color Kelbg
potential is finite at zero distance, thus removing
in a natural way the classical divergences and making any artificial cut-offs, often applied
(see, e.g., Ref. \cite{shuryak1}),  obsolete.
This color potential is a straightforward generalization of the corresponding potential of electromagnetic Coulomb plasmas
\cite{afilinov_pre04}.
The off-diagonal color Kelbg potential 
can be approximated  by the diagonal ones by means of
$\Phi_{12}({\bf r}_{12},{\bf r'}_{12},Q_1,Q_2)\approx
[\Phi_{12}({\bf r}_{12},{\bf r}_{12},Q_1,Q_2)
+\Phi_{12}({\bf r'}_{12},{\bf r'}_{12},Q_1,Q_2)]/2$.

Unfortunately such  rigorous consideration of multiparticle density matrix for particles interacting by potentials unbounded below  is not available. Therefore, following the experience gained in electromagnetic Coulomb plasmas, we use the following widely used ansatz \cite{feynm,zamalin},
which generalizes Eq. (\ref{GPERT}):
\begin{widetext}
\begin{eqnarray}
\rho^{(l)}
=
\rho^{(l)}_{0}\, \exp\left[-\Delta\beta
\frac{1}{2} \sum^N_{i,j (i\neq j)}
\Phi_{ij}
\left({\bf r}_i^{(l-1)}-{\bf r}_j^{(l-1)},{\bf r}_i^{(l)}-{\bf r}_j^{(l)}, Q_i,Q_j\right)
\right].
\label{ansatz}
\end{eqnarray}

Now we are able to construct ${R}$ of Eq. (\ref{Grho-pimc}).
The density matrix of noninteracting particles is known to be expressed
in terms of determinants and permanents of single-particle density matrices in the standard way.
{
These determinants and permanents take their origin from the (anti)symmetrization
discussed in Eq. (\ref{Grho-pimc}). }
Generalizing the electrodynamic plasma results \cite{filinov_ppcf01} to the quark-gluon plasma case,
we write approximate ${R}$
\begin{eqnarray}
&&
{R}\left(r^{(0)},r^{(1)}, \dots r^{(n+1)};Q; \{N\};\beta\right)=
\nonumber\\&=&
\exp\{-\beta \,U\}
 \,
\sum_{\sigma}
\left[
\prod\limits_{l=1}^n \prod\limits_{i=1}^N
\phi_{ii}\left( {\bf r}_{i}^{(l-1)},{\bf r}_{i}^{(l)},\Delta\beta \right) \,
\right]
\nonumber\\&\times&
\frac{{\rm per}\,\left\|\phi\left(r^{(n)},r^{(0)},\Delta\beta \right)\right\|_{N_g}
}{{\Lambda}_g^{3(n+1)N_g}(\Delta\beta)}
 \,
\frac{{\det}\,\left\|\phi\left(r^{(n)},r^{(0)},\Delta\beta \right)\right\|_{N_q}
\,}{{\Lambda}_q^{3(n+1)N_q}(\Delta\beta)}
\,
\frac{{\det}\,\left\|\phi\left(r^{(n)},r^{(0)},\Delta\beta \right)\right\|_{\overline{N}_q} \,}{{\Lambda}_{{\overline{q}}}^{3(n+1)\overline{N}_q}(\Delta\beta)}
\, ,
\label{Grho_s}
\end{eqnarray}
In Eq.~(\ref{Grho_s}) the effective total color interaction energy is
\begin{eqnarray}
U =
\frac{1}{n+1}\sum_{l=1}^{n+1} \frac{1}{2}
\sum_{i,j (i\neq j)}^N
\Phi_{ij}\left({\bf r}_i^{(l-1)}-{\bf r}_j^{(l-1)},{\bf r}_i^{(l)}-{\bf r}_j^{(l)},Q_i,Q_j\right).
\label{up}
\end{eqnarray}
\end{widetext}
Other quantities in  Eq.~(\ref{Grho_s}) are defined as follows:
\begin{eqnarray}
{\Lambda}^3_a(\beta)=\lambda^3_a\sqrt{\pi/2(\beta m_a)^5}
\label{Lambda}
\end{eqnarray}
with  $\lambda_a=\sqrt{2 \pi \beta / m_a}$
 being a thermal  wavelength of an $a$ type quasiparticle ($a=q, \overline{q},g$).
The antisymmetrization and symmetrization are taken into account by the symbols ``det'' and ``per''
denoting the determinant and permanent, respectively.
Eq.~(\ref{Grho_s}) is exact in the limit of $n\longrightarrow \infty$.
Indeed, since each factor in Eq. (\ref{Grho-pimc}) has an error of order of $1/(n+1)^\theta$ with $\theta > 1 $, the error of the whole product in the limit of $n\longrightarrow \infty$ equals zero.
Matrix $\phi\left(r,r',\Delta\beta \right)$ is defined by its matrix elements
\begin{eqnarray}
\label{phi_ii}
&&\phi_{ij} \left( {r},{r}',\Delta\beta \right)
= \frac{K_2 \left[ z_{ij}\left({ r},{r}',\Delta\beta \right)\right]}%
{\left[ z_{ij} \left({r},{r}',\Delta\beta\right) \right]^2}
\cr
&\times&
\delta_{}\,
\left[
\delta_{a_i,g} +
\left(
\delta_{a_i,q} + \delta_{a_i,\overline{q}}
\right)
\delta_{f_i,f_j}\,\delta_{\sigma_i,\sigma_j}
\right]
\end{eqnarray}
with
\begin{eqnarray}
z_{ij} \left( {r},{r}',\Delta\beta\right)
=\Delta\beta\, m_i\, \sqrt{1+\left|{\bf r}_{i}-{\bf r}'_{j}\right|^2/\Delta\beta^2}
\label{z_ii}
\end{eqnarray}
defined in terms of the modified Bessel function $K_2$. These matrix elements are nonzero
only for particles of the same type, i.e. $a_i=a_j$. Additional Kronecker symbols
in spin,  $\sigma_i$, and flavor,  $f_i$, indices of the particles are applicable
only to quark and antiquark  matrix elements.
They prevent  Pauli blocking for particles with different  spins and flavors.
The quantity $\phi$ describes the {\em relativistic measure} of trajectories
in the color path integral. This measure is associated with relativistic operator of kinetic energy in Eq. (\ref{Coulomb}).
In the limit of large  mass this  measure coincides with the Gaussian one used in Feynman-Wiener path integrals.
Due to factors $\delta_{a_i,a_j}\,\delta_{f_i,f_j}$ the matrix $\phi$ has a block structure corresponding
to different types of particles and different flavors of quarks and antiquarks. Subscripts $N_a$ near $\det$ and
${\rm per}$ operations precisely refer to the corresponding blocks, which in case of quarks and antiquarks
are still subdivided into sub-blocks related to flavors.

The dominant contribution to the partition function comes from
configurations in which the ``size '' of the quasiparticle cloud of 'beads' is of the
order of the Compton wavelength $\lambda_C=1/m_i$.
Thus, this path integral representation takes into account quantum uncertainty of the quasiparticle position.
In the limit of a large mass the spatial quasiparticle extension becomes much smaller than the average
interparticle distance. This makes possible an analytical integration over the 'beads' positions by the method of
steepest decent. As result the partition function is reduced to its classical limit involving point-like quasiparticles.

{
In fact, in Monte-Carlo simulations the pressure of the system is computed.
To obtained expression for the pressure we change the variable in Eq. (\ref{p_gen})
\begin{eqnarray}
\label{p_gen1}
P&=& \frac{\partial T\ln Z}{\partial V}  
= T\left[\frac{\alpha}{3V}\frac{\partial \ln Z  
 }{\partial \alpha}\right]_{\alpha=1},
\end{eqnarray}
where $\alpha=L/L_0$ ($V=\alpha^3 L_0^3$) is the length scaling parameter introduced
in physical quasiparticle coordinates. Details of derivation of
the final Monte-Carlo pressure estimator and
final intricate formula for path integral representation of partition function is presented
and discussed in \cite{zamalin,fusion,Fehske}.
Notice that only the maximal $\{N\}$ term in the
sum of Eq. (\ref{Gq-def}) is dominant in the
thermodynamic limit of the box volume $V \to \infty$.
Therefore, keeping only this maximal term,
corresponding to the canonical ensemble with $\{N\}$ numbers of particles,
 in the sum (\ref{Gq-def}), we arrive at the following expression for the
Monte-Carlo pressure estimator
\begin{widetext}
\begin{eqnarray}
&&
\frac{P}{P_0} = 1 - \frac{1}{N}
\frac{\left[3Z( \{N\},V,\beta )\right]^{-1}}{\, {{\Lambda}_g^{3(n+1)N_g}(\Delta\beta)} {{\Lambda}_q^{3(n+1)N_q}(\Delta\beta)}
{{\Lambda}_{{\overline{q}}}^{3(n+1)\overline{N}_q}(\Delta\beta)} }
\nonumber\\&\times&
 \int\limits d\mu Q\int dr^{(0)} dr^{(1)}\dots dr^{(n+1)}
{R}\left(r^{(0)},r^{(1)}, \dots ,r^{(n+1)};Q;\{N\};\beta\right)\;\delta\left(r^{(n+1)}- r^{(0)}\right)
\nonumber\\&\times&
\Bigg\{
\sum_{l=1}^{n} \sum_{i,j (i \neq j)}^{N}
\frac{\left( {\bf r}^l_{ij} \cdot {\bf r}_{ij}^0 \right)}{|{\bf r}^l_{ij}|}
\frac{\partial \beta U}{\partial |{\bf r}^l_{ij}|}
\nonumber\\&-&
\frac{\alpha}{
{\rm per}\,\left\|\phi\left(r^{(n)},r^{(0)},\Delta\beta \right)\right\|_{N_g} \, \cdot
{{\det}\,\left\|\phi\left(r^{(n)},r^{(0)},\Delta\beta \right)\right\|_{N_q} \,}
\, \cdot
{{\det}\,\left\|\phi\left(r^{(n)},r^{(0)},\Delta\beta \right)\right\|_{\overline{N}_q} \,}
}
\nonumber\\&\times&
\left[
\frac{{\partial }\,{\rm per}\,\left\|\phi\left(r^{(n)},r^{(0)},\Delta\beta \right)\right\|_{N_g} \, \cdot
{{\det}\,\left\|\phi\left(r^{(n)},r^{(0)},\Delta\beta \right)\right\|_{N_q} \,}\cdot
{{\det}\,\left\|\phi\left(r^{(n)},r^{(0)},\Delta\beta \right)\right\|_{\overline{N}_q} \,}
}
{\partial \alpha} \right]_{\alpha=1} \Bigg\}
\label{eos}
\end{eqnarray}
\end{widetext}
where $P_0$ is the pressure of the the ideal gas of quasiparticles,
${\bf r}^l_{ij}={\bf r}^l_{i}-{\bf r}^l_{j}$  is the distance between
beads with number $l$ of quasiparticles with numbers $i$ and $j$.}

{
The structure of Eq.~(\ref{eos}) is obvious. We have separated the classical ideal gas part
(first term). The ideal quantum part in excess of the classical one and the correlation
contributions are contained in the integral term.
The last term in curly brackets in Eq.~(\ref{eos}) is due to 
explicit
volume dependence of the exchange matrix. The main
advantage of Eqs.~(\ref{eos}) is that the explicit sum over permutations has been converted
into the determinant which can be computed very efficiently using standard linear
algebra methods.
Note that Eqs.~(\ref{eos}) contain the important
limit of an ideal quantum plasma in a natural way.
}

%
\section{Wigner Dynamics}\label{Wigner}

We are going to use Wigner formulation of quantum mechanics
for consideration of QGP kinetic properties.
Let us review the underlying ideas of the Wigner dynamics  for the simplest case, i.e.
for nonrelativistic colorless system of particles \cite{tatr1}.
The basis of our consideration is the Wigner representation of the
von Neumann equation -- a Wigner-Liouville equation (WLE).
To derive the WLE  for the simplest density matrix
$\rho(r,r^\prime,t)=\Psi(r,t)\Psi^{*}(r^\prime,t)$ of the $3D$  N-particle system
with $\Psi$ being an eigenfunction of a
Hamiltonian operator $\hat{H}=\sum^N_{i=1} \hat{p}^2_i/m+U$,
we introduce center-of-mass and relative coordinates
in a standard manner: $q \equiv (r + r^\prime )/2$
and $\xi \equiv r' - r$. Note that all these quantities
are $3N$-dimensional vectors. A Wigner distribution function (WF) is defined as
\bea
w\left(p,q,t\right) =
\frac{1}{(2\pi )^{6N}} \int
\rho \left(
   q+\frac{\xi}{2},q-\frac{\xi}{2},t
     \right) e^{i p \xi }\, d\xi.
\label{s1}
\eea
Here and below products of vector quantities like $p \xi$ are understood as
scalar products of $3N$ dimensional vectors.
Using this definition, one can derive the WLE for $w\left(p,q,t\right)$ \cite{tatr1,filmd, CicFil}.
Applying time derivative to definition (\ref{s1})  and taking into account that
\bea
i\frac{\partial}{\partial t}\Psi(r,t) =\hat{H} \Psi(r,t), \quad
i\frac{\partial}{\partial t}\Psi^{*}(r,t) =-\hat{H} \Psi^{*}(r,t)
\label{sh1}
\eea
we arrive at
\begin{widetext}
\bea
\frac{\partial w\left(p,q,t\right)}{\partial t} &=&
\frac{1}{(2\pi )^{6N}} \int d\xi \exp(i p \xi)
\frac{1}{i}\left[
\hat{H}(r,t) - \hat{H}(r',t)
\right] \rho(r,r^\prime,t)
\cr
&=&
\frac{1}{(2\pi )^{6N}} \int d\xi \exp(i p \xi)
\left\{
-\frac{i}{m}\frac{\partial^2}{\partial q \partial \xi}
+\frac{1}{i}
\left[U\left(q-\frac{\xi}{2}\right)-U\left(q+\frac{\xi}{2}\right)\right]
\right\}
\rho \left(q+\frac{\xi}{2},q-\frac{\xi}{2},t\right) \, .
\label{s2}
\eea
By means of integration by parts the first term in the braces can be transformed as follows
\bea
\frac{1}{(2\pi )^{6N}} \int d\xi \exp(i p \xi)
\left(-\frac{i}{m}\right)
\frac{\partial^2}{\partial q \partial \xi}
\rho \left(q+\frac{\xi}{2},q-\frac{\xi}{2},t\right)
=
-\frac{p}{m} \frac{\partial w\left(p,q,t\right)}{\partial q}
\label{s3}
\eea
while for the second one we obtain the following expression
\bea
&&
\frac{1}{(2\pi )^{6N}} \int d\xi \exp(i p \xi)
\frac{1}{i}
\left[U\left(q-\frac{\xi}{2}\right)-U\left(q+\frac{\xi}{2}\right)\right]
\rho \left(q+\frac{\xi}{2},q-\frac{\xi}{2},t\right)
\cr
&=&
\frac {4}{\left(2\pi \right)^{6N}}
\int ds \,w\left(p-s,q,t\right)
\int d{q'} \,
U\left(q-q'\right) \sin \left(2sq'\right)
\label{s4}
\eea
\end{widetext}
which results from substitution of  $\rho$ expressed in the form of inverse transformation to formula (\ref{s1}).

This way we arrive at the following form of the WLE
\bea
\hspace*{-5mm}
\frac{\partial w}{\partial t} +
\frac{p}{m}
\frac{\partial w}{\partial q}
-
\frac{\partial U(q)}{\partial q}
\frac{\partial w}{\partial p}
=
\int ds \,w\left(
p-s,q,t\right) \omega \left(s,q\right)
\label{s5}
\eea
where
\bea
\omega \left(s,q \right) =
&-&
\frac{\partial U(q)}{\partial q}\frac{d\delta \left(s\right) }{ds} 
\cr
&+&
\frac {4}{\left(2\pi \right)^{6N}}\int d{q'} \,
U\left(q-q'\right) \sin \left(2sq'\right)
\label{omega}
\eea
In the classical limit, $\hbar \rightarrow 0$,
the r.h.s. of Eq. (\ref{s5}) disappears and  Eq.~(\ref{s5}) is reduced
to the classical Liouville equation. This is the reason why we extracted the
term $\partial U(q)/\partial q$ from the r.h.s. of Eq. (\ref{s5}).

\subsection{Wigner Dynamics of Color Particles}\label{Color-Wigner}

Let us consider dynamics of QGP quasiparticles additionally characterized by
color variables $Q$ and derive the color WLE
for a density matrix $\rho(r,r^\prime,Q,t)$
of the $3D$  N-particle system,
 where, as before,
$Q$ denotes color degrees of freedom of all quarks, antiquarks and gluons.
Since color charges $Q$ are treated classically,
we consider only diagonal density matrix with respect to colors. Indeed, the $Q$ variable already
includes both canonical coordinate and momentum corresponding to classical color dynamics
(see Appendix I). Therefore, the density matrix takes the form
\bea
\rho(r,r^\prime,Q,t)=\underline{\rho}(r,r^\prime,Q,t)\prod_i \delta \left( Q_i-Q_i(t)\right)
\label{c-den-matrix}
\eea
where the product runs over all particles in the system and $Q_i(t)$ is a solution
of the classical equation of motion for color (\ref{Wong-eom-Q}).
Here $\underline{\rho}(r,r^\prime,Q,t)=\Psi(r,Q,t)\Psi^{*}(r^\prime,Q,t)$
is a quantum part of the density matrix
with $\Psi$ being an eigenfunction of the
Hamiltonian operator
described by Eq.~(\ref{Coulomb}) and $Q$ are already fixed c-numbers.

Now the definition of the corresponding WF reads
\bea
&&w\left(p,q,Q,t\right) 
\cr
&=&
\frac 1{(2\pi )^{3N}} \int
\rho \left(
   q-\frac{\xi}{2},q+\frac{\xi}{2},Q,t\right) e^{i p \xi }\, d\xi.
\label{ss1}
\eea
The quasiparticles are also characterized by spin and flavor,
which we do not explicitly include in the list of  quasiparticle degrees of freedom.
Notice that color degrees of freedom are also in the Wigner representation,
since $Q$ includes both color canonical coordinated and momenta, see Appendix I.
Now the WLE is defined by equation of the form \cite{ColWig11,ChoZahed}:
\bea
\frac{\partial w}{\partial t} &+& v  \frac{\partial w}{\partial q}  +
F  \frac{\partial w}{\partial p} +
\Upsilon  \frac{\partial w}{\partial Q} 
\cr
&=&
\int ds \,w\left(
p-s,q,Q,t\right) \omega \left(s,q\right),
\label{sq5}
\eea
where $v=\{{\bf v}_i\}$ is 3N-dimensional vector of velocities of all quasiparticles, cf.
Eq. (\ref{Wong-eom-r}),
$F = -\partial U^C(q,Q)/\partial q$ is a set of the color-electric forces experienced by  all quasiparticles,
cf. Eq. (\ref{Wong-eom-p}),
$\Upsilon=\{\Upsilon^a_i\}$ is an 8N-dimensional vector of
driving forces in Wong's equation of motion for the color charge (\ref{Wong-eom-Q}), and
$\omega \left(s,q \right)$ is
defined by Eq. (\ref{omega}).

The classical part of WLE (\ref{sq5}), i.e. the l.h.s. of it, can be easily derived
e.g. from the Wong's equations of motion (\ref{Wong-eom-p})-(\ref{Wong-eom-Q})
for the color-charged particles (see Ref. \cite{ChoZahed}).
In particular, the term $\Upsilon  {\partial w}/{\partial Q}$ natirally results from
$(dQ(t)/dt)  {\partial w}/{\partial Q}$ and Wong's equation (\ref{Wong-eom-Q}).
Since we confine ourselves to classical dynamics of color, we do not need any further
(quantum) consideration for it. The quantum space dynamics [i.e. the r.h.s. of Eq. (\ref{sq5})]
is derived completely in the same way as it was described above, see Eqs. (\ref{s2})-(\ref{s5}),
with minor complications due to relativistic kinematics.

\subsection{Wigner representation of time correlation functions}\label{s:W-representation}

In computations of transport properties, like viscosity,
our starting point is the general Kubo expression for the
canonical ensemble-averaged operator
\cite{zubar}
\begin{equation}
\breve{C}_{BA}(t) =Z^{-1}\mbox{Tr}\left\{e^{-\beta{H}} \hat{B}\,e^{i\hat{H}t}\,
\hat{A}\,e^{-i\hat{H}t} \right\},
\label{GFA1}
\end{equation}
where $\hat{B}$ and $\hat{A}$ are quantum operators of dynamic
quantities under consideration and
$Z\left( N,V,T \right) =\mbox{Tr}\left\{e^{-\beta\hat{H}}\right\}$ is the
canonical partition function. Frequently a symmetric time-correlation function
is also used \cite{doll}:
\begin{equation}
C_{BA}(t) =Z^{-1}\mbox{Tr}\left\{\hat{B}\,e^{i\hat{H}t_c^{*}}\,
\hat{A}\,e^{-i\hat{H}t_c} \right\},
\label{CFA1}
\end{equation}
where $t_c=t-i \beta /2$ is  a complex-valued quantity
including the inverse temperature $\beta =1/T$.
 The Fourier transforms of  $\breve{C}_{BA}(t)$ and $C_{BA}(t)$
are  related as \cite{doll}
\begin{eqnarray}
C_{BA}(\omega) =\exp\left(- \frac{\beta \omega}{2}\right)\breve{C}_{BA}(\omega).
\label{GF}
\end{eqnarray}
As a consequence, transport coefficients described by zero-frequency ($\omega=0$) Fourier components
can be obtained from the symmetric time-correlation functions, which may offer certain
computational advantages. This symmetric form is used below.

The Wigner representation of the time-correlation function in a
$6N$-dimensional space can be written as
\begin{eqnarray}
&&C_{BA}(t)=(2\pi)^{-6N}
\label{CFA}
\\
&\times&
\int d\overline{pq\mu Q} d\widetilde{pq\mu Q}
\,B(\overline{pqQ})A(\widetilde{pqQ})\,
W\left(\overline{pqQ};\widetilde{pqQ} ;t;\beta \right)
\nonumber
\end{eqnarray}
where  we introduced a short-hand notation
for phase space points in (6N+8N)-dimesional space, $\overline{pqQ}$ and $\widetilde{pqQ}$, with
$p$, $q$ and $Q$ comprising the momenta, coordinates  and
color variables, respectively, of all particles of the system.
Here $A(pqQ)$ denotes the Weyl's symbol \cite{tatr1} of the operator $\hat{A}$:
\begin{equation}
A(\widetilde{pqQ})=\int d\widetilde{\xi} \, \exp(-i \widetilde{p}\widetilde{\xi})
\left\langle \widetilde{q}-\frac{\widetilde{\xi}}{2},\widetilde{Q}\left |\hat{A}\right|
\widetilde{q}+\frac{\widetilde{\xi}}{2},\widetilde{Q}\right\rangle,
\label{WS}
\end{equation}
 and similarly for the operator
$\hat{B}$, while
$W\left(\overline{pqQ};\widetilde{pqQ} ;t;\beta \right)$ is the spectral density expressed as
\begin{widetext}
\begin{eqnarray}
W\left(\overline{pqQ};\widetilde{pqQ} ;t;\beta \right) = Z^{-1}
\sum_{\widetilde{\sigma},\overline{\sigma}}
\int\int d\overline{\xi}d\widetilde{\xi} \,
e^{i\overline{p} \overline{\xi}}e^{i\widetilde{p}\widetilde{\xi}}
\left\langle \overline{q}+\frac{\overline{\xi}}{2},\overline{Q}\left|e^{i\hat{H}t_c^{*}}\right|
  \widetilde{q}-\frac{\widetilde{\xi}}{2},\widetilde{Q} \right\rangle 
    \left\langle \widetilde{q}+\frac{\widetilde{\xi}}{2},\widetilde{Q}\left|e^{-i\hat{H}t_c}\right|
   \overline{q}-\frac{\overline{\xi}}{2},\overline{Q}\right\rangle. 
\label{inw}
\end{eqnarray}
In Eq. (\ref{CFA}) we silently assumed that operators $\hat{A}$ and $\hat{B}$ do not depend
on spin variables. Therefore, summation over spins $\widetilde{\sigma}$ and $\overline{\sigma}$
can be safely moved to the definition of $W$. Here and below we do not explicitly
write spin variables, if they are not essential.
The time-correlation function $C_{BA}(t)$
is a linear functional of the spectral density $W$. Thus, the problem of its treatment
is reduced to the consideration of
the spectral density evolution.

As it follows from Eqs.~(\ref{sq5}) and Ref. \cite{filmd1},
the following system of the WL integro-differential equations describe
the time evolution of the color spectral density $W$:
\bea
\frac{\partial W}{\partial t} +
\overline{v}
\frac{\partial W}{\partial \overline{q}}
+
F
\frac{\partial W}{\partial \overline{p}}
+
\Upsilon  \frac{\partial W}{\partial \overline{Q}}
&=&
\int ds \,W\left(
\overline{p}-s,\overline{qQ};\widetilde{pqQ};t;\beta\right) \omega \left(s,\overline{q},\right),
\label{ur1}
\\
-\frac{\partial W}{\partial t} +
\widetilde{v}
\frac{\partial W}{\partial \widetilde{q}}
+
F
\frac{\partial W}{\partial \widetilde{p}}
+
\Upsilon  \frac{\partial W}{\partial \widetilde{Q}}
&=&
\int ds \,W\left(\overline{pqQ};
\widetilde{p}-s,\widetilde{qQ};t;\beta\right) \omega \left(s,\widetilde{q}\right),
\label{ur2}
\eea
 where as before $\omega \left(s,q \right)$
is defined by Eq. (\ref{omega}).
This equations are derived precisely
in the same way as those of Eqs. (\ref{sq5}) and (\ref{s2})-(\ref{s5}),
only the Hamiltonian $H$ appears here as a result of time derivation of
exponent functions, $e^{i\hat{H}t}$ and $e^{-i\hat{H}t}$, rather than
from application of equations of motion (\ref{sh1}) in Eq. (\ref{s2}).
Notice that while Eq. (\ref{ur1}) describes evolution in the positive
time direction, Eq. (\ref{ur2}) specifies propagation in the reverse
time direction. This happens because of the presence the direct time
$e^{-i\hat{H}t}$ and reverse time $e^{i\hat{H}t}$ evolution operators
in definition of the time-correlation function (\ref{CFA1}).

Now using Eqs. (\ref{ur1}) and (\ref{ur2}) we can obtain an
integral equation \cite{tatr1,filmd,CicFil,filmd1} for
$W\left(\overline{pqQ};\widetilde{pqQ} ;t;\beta \right)$
\begin{eqnarray}
&&
W\left(\overline{pqQ};\widetilde{pqQ} ;t;\beta \right)
\nonumber\\&&
=
\left\{
\int d\overline{p_0q_0\mu Q_0} \, d\widetilde{p_0q_0\mu Q_0} \,
G\left(\overline{pqQ},\widetilde{pqQ},t;\overline{p_0q_0Q_0},\widetilde{p_0q_0Q_0},0\right)
\, W(\overline{p_0q_0Q_0};\widetilde{p_0q_0Q_0};t=0, \beta )
\right.
\nonumber\\&&
+\frac{1}{2}
\left(
\int_0^t dt' \int ds \,
\int d\overline{p'q'\mu Q'} \, d\widetilde{p'q'\mu Q'} \,
G \left(\overline{pqQ},\widetilde{pqQ},t;\overline{p'q'Q'},\widetilde{p'q'Q'},t'\right) \,
\right.
 \nonumber\\&&\times
\left.\left.
\left[W(\overline{p}'-s,\overline{q'Q'};\widetilde{p'q'Q'};
   t' ;\beta) \, \omega (s,\overline{q}')-
W(\overline{p'q'Q'};\widetilde{p}'-s,\widetilde{q'Q'};
   t' ;\beta) \,\omega (s,\widetilde{q}')\right]
\vphantom{\int}\right) \right\},
\label{w}
\end{eqnarray}
with Green function
\bea
&&G \left(\overline{pqQ},\widetilde{pqQ},t;\overline{p'q'Q'},\widetilde{p'q'Q'},t'\right)
\nonumber\\&=&
\delta\left(\overline{p}-\overline{p}(t;\overline{p'q'Q'},t')\right)
\delta \left(\overline{q}-\overline{q}(t;\overline{p'q'Q'},t')\right)
\delta \left(\overline{Q}-\overline{Q}(t;\overline{p'q'Q'},t')\right)
\nonumber\\&\times&
\delta \left(\widetilde{p}-\widetilde{p}(t;\widetilde{p'q'Q'},t')\right)
\delta\left(\widetilde{q}-\widetilde{q}(t;\widetilde{p'q'Q'},t')\right)
\delta \left(\widetilde{Q}-\widetilde{Q}(t;\widetilde{p'q'Q'},t')\right).
\label{greenk}
\eea
\end{widetext}
describing propagation of the spectral density along classical trajectories  in positive time direction
\bea 
\frac{d\overline{p}(t;\overline{p'q'Q'},t')}{dt}&=&
\mbox{\ \ }\frac{1}{2}F(\overline{qQ}_t) , 
\cr
\frac{d\overline{q}(t;\overline{p'q'Q'},t')}{dt}&=&
\mbox{\ \ }\frac{1}{2} \overline{v}[\overline{p}(t;\overline{p'q'Q'},t')],  
\cr
\frac{d\overline{Q}(t;\overline{p'q'Q'},t')}{dt}&=&
\mbox{\ \ }\frac{1}{2}\Upsilon(\overline{qQ}_t), 
\label{HW-dir}
\eea
and in the reverse time direction
\bea
\frac{d\widetilde{p}(t;\widetilde{p'q'Q'},t')}{dt}&=&-\frac{1}{2}F(\widetilde{qQ}_t),
\cr
\frac{d\widetilde{q}(t;\widetilde{p'q'Q'},t')}{dt}&=&-\frac{1}{2}\widetilde{v}[\widetilde{p}(t;\widetilde{p'q'Q'},t')],
\cr
\frac{d\widetilde{Q}(t;\widetilde{p'q'Q'},t')}{dt}&=&-\frac{1}{2}\Upsilon(\widetilde{qQ}_t),
\label{HW-inv}
\eea   
where
$(\widetilde{qQ}_t)=[\widetilde{q}(t;\widetilde{p'q'Q'},t'),\widetilde{Q}(t;\widetilde{p'q'Q'},t')]$
and similarly for bared quantities.
These equations of motion are supplemented by initial conditions at time $t=0$
\bea
\overline{p}(t;\overline{p_0q_0Q_0},0)&=&\overline{p_0},
\cr
\overline{q}(t;\overline{p_0q_0Q_0},0)&=&\overline{q_0},
\cr
\overline{Q}(t;\overline{p_0q_0Q_0},0)&=&\overline{Q_0},
\label{ini-dir}
\\
\widetilde{p}(t;\widetilde{p_0q_0Q_0},0)&=&\widetilde{p_0},
\cr
\widetilde{q}(t;\widetilde{p_0q_0Q_0},0)&=&\widetilde{q_0},
\cr
\widetilde{Q}(t;\widetilde{p_0q_0Q_0},0)&=&\widetilde{Q_0}.
\label{ini-dir0}
\eea
and by initial conditions at time $t=t'$
\bea
\overline{p}(t';\overline{p'q'Q'},t')&=&\overline{p}',\quad
\cr
\overline{q}(t';\overline{p'q'Q'},t')&=&\overline{q}',\quad
\cr
\overline{Q}(t';\overline{p'q'Q'},t')&=&\overline{Q}',
\label{ini-inv}
\\
\widetilde{p}(t';\widetilde{p'q'Q'},t')&=&\widetilde{p}',\quad
\cr
\widetilde{q}(t';\widetilde{p'q'Q'},t')&=&\widetilde{q}',\quad
\cr
\widetilde{Q}(t';\widetilde{p'q'Q'},t')&=&\widetilde{Q}'.
\label{ini-inv0}
\eea
In fact, Eqs. (\ref{HW-dir}) are Wong's equations of motion but written for half-time ($t/2$).
Similarly, Eqs. (\ref{HW-inv}) are half-time Wong's e
quations of motion reversed in time.
This happens because the time correlation is taken between instants in the past and the future
with the initial conditions fixed in between these instants, i.e. at $t=0$ the spectral density
$W(\overline{pqQ}_0;\widetilde{pqQ}_0;t=0,\beta)=W_0(\overline{pqQ}_0;\widetilde{pqQ}_0;\beta)$
is fixed as it is described in the next subsection.
{ 
The right-hand sides
of  equations  (\ref{HW-dir}) and (\ref{HW-inv}) include
interparticle interaction that can be arbitrary strong.}

{
Solution of the integral equation (\ref{w}) can be obtained in a form
of iterative series with absolute convergence.}
In this work, we take into account only the first term of this iteration series:
\begin{eqnarray}
&&
W\left(\overline{pqQ};\widetilde{pqQ} ;t;\beta \right)
\simeq 
\int d\overline{p_0q_0\mu Q_0} \, d\widetilde{p_0q_0\mu Q_0} \,
\cr
&\times&
G\left(\overline{pqQ},\widetilde{pqQ},t;\overline{p_0q_0Q_0},\widetilde{p_0q_0Q_0},0\right)
\, 
\cr
&\times&
W_0(\overline{pqQ}_0;\widetilde{pqQ}_0;\beta ).
\label{1st-w}
\end{eqnarray}
Notice that if the initial
$W_0(\overline{pqQ}_0;\widetilde{pqQ}_0;\beta)$
is chosen appropriately \cite{tatr1}, i.e. such that it contains all powers
of the Planck's constant, then the first term of the iterative series, i.e.  Eq.~(\ref{1st-w}),
describes propagation of
a {\em quantum} initial spectral density along classical trajectories. Other (higher) terms
describe propagation of the initial spectral density along the analogous trajectories
but perturbated by momentum jumps resulted from the convolution structure of
integral term in Eq.~(\ref{w}). From physical point of view these jumps
relate to quantum ($[p,q]$) uncertainty (for details see discussion in Ref.~\cite{filmd}).
As it was found in Refs.
\cite{filmd, CicFil,filmd6}, the main contribution to WF
comes from the trajectories without jumps, i.e. from the first term of the iterative series,
if the motion takes place in a classically accessible region, which is the case here.
Nevertheless calculations with higher-order
iterative terms are in progress and will be reported elsewhere.

{
As known, the classical limit for multi-component Coulomb system does not exist since
the stability of Coulomb systems is only provided by quantum effects.
That was the physical reason for  arising the color Kelbg potential
in  the patition function and Eq. (\ref{ppsi}) for initial
condition (see below).
To take into account this quantum effect 
we replace the color Coulomb potential
$U^C$  by the Kelbg one $\Phi$ in quantities $F$ and $\Upsilon$ 
defining  propagation of the
spectral density along the classical trajectories, see Eqs. (\ref{HW-dir}) and (\ref{HW-inv}).
By this replacement we are able to take into
account certain higher-order quantum terms of the iteration series
presenting the solution of integral equation (\ref{w}).
From the practical point of view, it allows us to avoid problems
due to singular character of the Coulomb potential.
}

\subsection{Initial conditions}\label{s:InCd}

The initial function $W_0$ is expressed in terms of
matrix elements of density matrix considered in  section \ref{s:pimc}.
Accordingly to Eq. (\ref{inw}) for $t=0$ and definition of the density matrix
$\rho$, cf. Eq. (\ref{Z-def}), we have
\begin{widetext}
\begin{eqnarray}
W_0\left(\overline{pqQ};\widetilde{pqQ} ;\beta \right) &=& Z^{-1}
\sum_{\widetilde{\sigma},\overline{\sigma}}
\int d\overline{\xi}d\widetilde{\xi} \,
e^{i\overline{p} \overline{\xi}}e^{i\widetilde{p}\widetilde{\xi}}
\left\langle \overline{q}+\frac{\overline{\xi}}{2},{\overline{Q}}\left|
\rho\left(\frac{\beta}{2}\right)  \right|
  \widetilde{q}-\frac{\widetilde{\xi}}{2},{\widetilde{Q}} \right\rangle 
    \left\langle \widetilde{q}+\frac{\widetilde{\xi}}{2},{\widetilde{Q}}\left|
   \rho\left(\frac{\beta}{2}\right) \right|
   \overline{q}-\frac{\overline{\xi}}{2},{\overline{Q}}\right\rangle
\cr
&\times&
\delta(\overline{Q}-\widetilde{Q}) \, \delta_{\widetilde{\sigma},\overline{\sigma}}
   .
\label{W0}
\end{eqnarray}
\end{widetext}
Thus, the problem is reduced to calculation of matrix elements of density matrix
$\rho$, which is similar to that we did in sect. \ref{Thermodynamics} devoted to
thermodynamics. Only now we need nondiagonal matrix elements rather than diagonal ones,
as in sect. \ref{Thermodynamics}.

As before, cf. Eq. (\ref{Grho-pimc}), let us subdivide  $\rho\left(\beta/2\right)$
into beards using the operator identity
$$e^{-\beta \hat{H}/2}= e^{-\Delta \beta' {\hat H}} \cdot
e^{-\Delta \beta' {\hat H}}\dots  e^{-\Delta \beta' {\hat H}},$$
where the r.h.s. contains $n+1$ identical factors with $\Delta \beta' = \beta/[2(n+1)]$,
so
\begin{eqnarray}
&&
\left\langle \overline{q}+\frac{\overline{\xi}}{2},{Q}\left|
\rho\left(\frac{\beta}{2}\right)  \right|
  \widetilde{q}-\frac{\widetilde{\xi}}{2},{Q} \right\rangle
\nonumber\\
&\simeq&
\int 
d\overline{r}^{(1)} \dots d\overline{r}^{(n)} \,
\overline{\rho}^{(1)}\,
\overline{\rho}^{(2)}\,\overline{\rho}^{(3)} \, \dots
\overline{\rho}^{(n)}\overline{\rho}^{(n+1)}
\label{Grho-pimc-W}
\end{eqnarray}
where $\overline{\rho}^{(l)}$ ($l=1,\dots, n+1$)
are  defined  by Eqs.
 (\ref{rho(l)}) and (\ref{ansatz}) with
$$
\overline{r}^{(0)}=\overline{q}+\frac{\overline{\xi}}{2},\quad
\overline{r}^{(n+1)}= \widetilde{q}-\frac{\widetilde{\xi}}{2}
$$
and  $\Delta \beta$ replaced by $\Delta \beta'$.
The bar sign above $\overline{\rho}^{(l)}$
means that these quantities depend on ``bared'' variables $\overline{r}^{(l)}$.
Similarly, after antisymmetrization
we obtain
\begin{eqnarray}
&&
\left\langle \widetilde{q}+\frac{\widetilde{\xi}}{2},{Q}\left|
\rho\left(\frac{\beta}{2}\right)  \right|
  \overline{q}-\frac{\overline{\xi}}{2},{Q} \right\rangle 
\nonumber\\
&\simeq&
\int 
d\widetilde{r}^{(1)}\dots
d\widetilde{r}^{(n+1)} \,
\widetilde{\rho}^{(1)}\,
\widetilde{\rho}^{(2)}\,\widetilde{\rho}^{(3)} \, \dots
\widetilde{\rho}^{(n)}\,
\nonumber\\&\times&
\left[\sum_{P_q} \sum_{P_{ \widetilde{q}}}\sum_{P_g}(- 1)^{\kappa_{P_q}+ \kappa_{P_{\overline{q}}}}
{\hat P_q} {\hat P_{ \widetilde{q}}}{\hat P_g}\widetilde{\rho}^{(n+1)}\right],
\, 
 \label{Grho-pimc-W1}
\end{eqnarray}
with
$$
\widetilde{r}^{(0)}=\widetilde{q}+\frac{\widetilde{\xi}}{2},\quad   
\widetilde{r}^{(n+1)}= \overline{q}-\frac{\overline{\xi}}{2}, 
$$
where the ``tilde'' functions $\widetilde{\rho}^{(l)}$ 
depend on ``tilde'' variables.
Similarly to Eq. (\ref{Grho-pimc}) it is enough to perform the 
symmetrization-antisymmetrization only in a single matrix element in Eq. (\ref{W0}).

{
In Sect. \ref{Thermodynamics} we used the approximate expression for elements
of the density matrix based on Eq. (\ref{ansatz}).
Here it is not practical.
An alternative approximation can be derived by means of expression (\ref{Mori}) for the
density matrix operator.
}
Notice that in fact $\exp\{-(\Delta\beta'/2){\hat K}\}$ is a density matrix
of the non-interacting system.
This symmetrized form results in the following approximation to the matrix elements
\begin{widetext}
\begin{eqnarray}
\rho^{(l)}
&=&
\rho\left(r^{(l-1)},r^{(l)},Q;\{N\};\Delta\beta'/2\right)
\cr
&=&
\int d q^{(l)} \;
\rho_0\left(r^{(l-1)},q^{(l)},Q;\{N\};\Delta\beta'/2\right)\;
\rho_0\left(q^{(l)},r^{(l)},Q;\{N\};\Delta\beta'/2\right)
\cr
&\times&
\exp\left[-\Delta\beta'
\frac{1}{2} \sum^N_{i,j (i\neq j)}
\Phi_{ij}
\left({\bf q}_i^{(l)}-{\bf q}_j^{(l)},{\bf q}_i^{(l)}-{\bf q}_j^{(l)}, Q_i,Q_j\right)
\right]
,
\label{rho(ll)}
\end{eqnarray}
\end{widetext}
where $\Phi_{ij}$ is the diagonal part of the color Kelbg potential, see Eq. (\ref{kelbg-d}).
This approximate form has the same accuracy as that in Eq.  (\ref{rho(l)} ).
At the same time it allows us to explicitly perform Fourier transforms in Eq. (\ref{W0}),
since the $\xi$ dependence now occurs only in the $\rho_0$ factors.

Thus, based on the above approximation for $\rho^{(l)}$ we are able to
explicitly evaluate integrals over $\overline{\xi}$ and $\widetilde{\xi}$
\begin{eqnarray}
\int d\widetilde{\xi} \,
e^{i\widetilde{p} \widetilde{\xi}}
\overline{\rho}^{(n+1)}
\widetilde{\rho}^{(1)}
&=&
\overline{\rho}^{(n+1)}
\widetilde{\rho}^{(1)}
\varphi  \left(\widetilde{p};\overline{r}^{(n)},\widetilde{r}^{(1)}\right)
\label{bar-xi-int}
\\
\int d\overline{\xi} \,
e^{i\overline{p}\overline{\xi}}
\widetilde{\rho}^{(n+1)}
\overline{\rho}^{(1)}
&=&
\widetilde{\rho}^{(n+1)}
\overline{\rho}^{(1)}
\varphi  \left(\overline{p};\widetilde{r}^{(n)},\overline{r}^{(1)}\right)
\label{tilde-xi-int}
\end{eqnarray}
where on the r.h.s. of these equations and below the marginal coordinates
take already the following values
$$
\overline{r}^{(0)}=\overline{q},\quad    
\overline{r}^{(n+1)}= \widetilde{q},   
$$
$$
\widetilde{r}^{(0)}=\widetilde{q},\quad   
\widetilde{r}^{(n+1)}= \overline{q}, 
$$
and the complex-valued function $\varphi$ is defined as
\bea  
&&
\varphi \left( p;r^{\prime },r^{\prime \prime }\right) 
\cr
&=& \prod\limits_{i=1}^N
\left(2\lambda'^2_i\right)^{3/2}
\exp \left[-\frac{1}{2\pi}\left( {\bf p}_i \lambda'_i
+i\pi \frac{{\bf r}_i^{\prime}-{\bf r}_i^{\prime \prime}}{\lambda'_i}
\right)^2 \right]
\hspace*{5mm}
\label{def-varphi}
\eea  
with $\lambda'_i=\sqrt{ \pi \Delta\beta' / m_i }$
being  the $i$-particle thermal wave lengths related to temperature $2/\Delta\beta'$.

Substituting these expressions into Eq. (\ref{W0}),
 we arrive at  \cite{filmd1,filmd6,feynm}
\begin{widetext}
\bea
W_0(\overline{pqQ};\widetilde{pqQ};\beta )
&=& \frac{1}{Z}
\int
d\overline{r}^{(1)}\dots
d\overline{r}^{(n)} \,
d\overline{q}^{(1)}\dots
d\overline{q}^{(n+1)} \,
d\widetilde{r}^{(1)}\dots
d\widetilde{r}^{(n)} \,
d\widetilde{q}^{(1)}\dots
d\widetilde{q}^{(n+1)} \,
\nonumber\\&\times&
\Psi \left(\overline{pqQ};\widetilde{pqQ};\overline{r}^{(1)}, \dots ,\overline{r}^{(n)};
\overline{q}^{(1)}, \dots ,\overline{q}^{(n+1)};
\widetilde{r}^{(1)},\dots,\widetilde{r}^{(n)};
\widetilde{q}^{(1)},\dots,\widetilde{q}^{(n+1)}; \beta \right)
\label{Psi-def-1}
\eea
with
\begin{eqnarray}
&&
\Psi \left(\overline{pqQ};\widetilde{pqQ};\overline{r}^{(1)}, \dots ,\overline{r}^{(n)};
\overline{q}^{(1)}, \dots ,\overline{q}^{(n+1)};
\widetilde{r}^{(1)},\dots,\widetilde{r}^{(n)};
\widetilde{q}^{(1)},\dots,\widetilde{q}^{(n+1)}; \beta \right)
\nonumber\\&=&
\exp\left\{-\beta \left(\overline{U}+\widetilde{U}\right)\right\}
\delta(\overline{Q}-\widetilde{Q})
\nonumber\\&\times&
\sum_{\widetilde{\sigma},\overline{\sigma}}
\left[
\prod_{l=1}^{n+1}
\rho_0\left(\overline{r}^{(l-1)},\overline{q}^{(l)},\overline{Q};\{N\};\Delta\beta'/2\right)\;
\rho_0\left(\overline{q}^{(l)},\overline{r}^{(l)},\overline{Q};\{N\};\Delta\beta'/2\right)
\right]
\varphi  \left(\widetilde{p};\overline{r}^{(n)},\widetilde{r}^{(1)}\right)
\nonumber\\&\times&
\left[
\prod_{l=1}^{n}
\rho_0\left(\widetilde{r}^{(l-1)},\widetilde{q}^{(l)},\widetilde{Q};\{N\};\Delta\beta'/2\right)\;
\rho_0\left(\widetilde{q}^{(l)},\widetilde{r}^{(l)},\widetilde{Q};\{N\};\Delta\beta'/2\right)
\right]
\varphi  \left(\overline{p};\widetilde{r}^{(n)},\overline{r}^{(1)}\right)
\nonumber\\&\times&
\left[
 \sum_{P_q} \sum_{P_{ \overline{q}}}\sum_{P_g}(- 1)^{\kappa_{P_q}+ \kappa_{P_{\overline{q}}}}
{\hat P_q} {\hat P_{ \overline{q}}}{\hat P_g}
\rho_0\left(\widetilde{q}^{(n+1)},\widetilde{r}^{(n+1)},\widetilde{Q};\{N\};\Delta\beta'/2\right)\;
\delta_{\overline{\sigma},\widetilde{\sigma}} \,
{\cal S}(\widetilde{\sigma}, {\hat P_q}{\hat P_{ \overline{q}}}{\hat P_g} \widetilde{\sigma}^{\prime})\big|_{\widetilde{\sigma}^{\prime}=\widetilde{\sigma} }
\right] \;
\label{ppsi}
\end{eqnarray}
where
\begin{eqnarray}
\overline{U} =
\frac{1}{2} \frac{1}{n+1}\sum_{l=1}^{n+1}
\sum_{i,j (i\neq j)}^N
\Phi_{ij}\left(\overline{\bf q}_i^{(l)}-\overline{\bf q}_j^{(l)},
\overline{\bf q}_i^{(l)}-\overline{\bf q}_j^{(l)},\overline{Q}_i,\overline{Q}_j\right).
\label{U-bar}
\\
\widetilde{U} =
\frac{1}{2} \frac{1}{n+1}\sum_{l=1}^{n+1}
\sum_{i,j (i\neq j)}^N
\Phi_{ij}\left(\widetilde{\bf q}_i^{(l)}-\widetilde{\bf q}_j^{(l)},
\widetilde{\bf q}_i^{(l)}-\widetilde{\bf q}_j^{(l)},\widetilde{Q}_i,\widetilde{Q}_j\right).
\label{U-tilde}
\end{eqnarray}
Applying the notation of Sect. II (cf. Eqs. (\ref{Lambda}) and Eq.
(\ref{phi_ii})), we finally arrive at
\begin{eqnarray}
&&
\Psi \left(\overline{pqQ};\widetilde{pqQ};{r}^{(1)}, \dots ,\overline{r}^{(n)};
\overline{q}^{(1)}, \dots ,\overline{q}^{(n+1)};
\widetilde{r}^{(1)},\dots,\widetilde{r}^{(n)};
\widetilde{q}^{(1)},\dots,\widetilde{q}^{(n+1)}; \beta \right)
\nonumber\\&=&
\exp\left\{-\beta \left(\overline{U}+\widetilde{U}\right)\right\}\,
\delta(\overline{Q}-\widetilde{Q})
\nonumber\\&\times&
\sum_{\widetilde{\sigma},\overline{\sigma}}
\left[
\prod_{l=1}^{n+1}
\left(
\prod\limits_{i=1}^N
\phi_{ii}\left(\overline{r}^{(l-1)},\overline{q}^{(l)},\Delta\beta'/2\right)\;
\right)
\left(
\prod\limits_{i=1}^N
\phi_{ii}\left(\overline{q}^{(l)},\overline{r}^{(l)},\Delta\beta'/2\right)
\right)
\right]
\varphi  \left(\widetilde{p};\overline{r}^{(n)},\widetilde{r}^{(1)}\right)
\nonumber\\&\times&
\left[
\prod_{l=1}^{n}
\left(
\prod\limits_{i=1}^N
\phi_{ii}\left(\widetilde{r}^{(l-1)},\widetilde{q}^{(l)},\Delta\beta'/2\right)\;
\right)
\left(
\prod\limits_{i=1}^N
\phi_{ii}\left(\widetilde{q}^{(l)},\widetilde{r}^{(l)},\Delta\beta'/2\right)
\right)
\right]
\nonumber\\&\times&
\frac{{\det}\,\left\|\phi
\left(\widetilde{q}^{(n+1)},\widetilde{r}^{(n+1)},\Delta\beta'/2\right)
\right\|_{N_q}
}{{\Lambda}_q^{6(n+1)N_q}(\Delta\beta'/2)}
\,
\frac{{\det}\,\left\|\phi
\left(\widetilde{q}^{(n+1)},\widetilde{r}^{(n+1)},\Delta\beta'/2\right)
\right\|_{\overline{N}_q}
}{{\Lambda}_{{\overline{q}}}^{6(n+1)\overline{N}_q}(\Delta\beta'/2)}
\nonumber\\&\times&
\frac{{\rm per}\,\left\|\phi
\left(\widetilde{q}^{(n+1)},\widetilde{r}^{(n+1)},\Delta\beta'/2\right)
\right\|_{N_g}
}{{\Lambda}_g^{6(n+1)N_g}(\Delta\beta'/2)}
\varphi  \left(\overline{p};\widetilde{r}^{(n)},\overline{r}^{(1)}\right)
 \,
\delta_{\overline{\sigma},\widetilde{\sigma}}
\nonumber\\&&
\label{ppsi1}
\end{eqnarray}
with
$$
\overline{r}^{(0)}=\overline{q},\quad
\overline{r}^{(n+1)}= \widetilde{q},\quad
\widetilde{r}^{(0)}=\widetilde{q},\quad
\widetilde{r}^{(n+1)}= \overline{q}.
$$

In the limit $n \rightarrow \infty$  this expression exactly gives the product of matrix elements in Eq. (\ref{W0}) in the form of path integrals multiplied
by a limiting expression of the $\varphi$    functions.
According to the Lebesque-Dirac delta theorem (see Appendix II), the
 $\overline{\rho}^{()}\widetilde{\rho}^{()}\varphi$  products
in integral  (\ref{Psi-def-1}) in the limit $n \to \infty$
are equivalent to the to the following real valued expressions
\begin{eqnarray}
\overline{\rho}^{(n+1)}\widetilde{\rho}^{(1)}
\varphi  \left(\widetilde{p};\overline{r}^{(n)},\widetilde{r}^{(1)}\right)
& \stackrel{n \to \infty}{\longrightarrow} &
\overline{\rho}^{(n+1)}\widetilde{\rho}^{(1)}
\prod\limits_{i=1}^N
\left(\left(2\lambda^2_i\right)^{3/2}
\exp \left[-\frac{\widetilde{\bf p}^2_i \lambda_i^2}{2\pi} \right] \right)
\left(\frac{\lambda'_i}{\pi}\right)^3
\delta\left(\overline{\bf r}_i^{(n)}-\widetilde{\bf r}_i^{(1)}\right)
\label{tilde-xi-intl-1}
\\
%
\widetilde{\rho}^{(n+1)}\overline{\rho}^{(1)}
\varphi  \left(\overline{p};\widetilde{r}^{(n)},\overline{r}^{(1)}\right)
&\stackrel{n \to \infty}{\longrightarrow} &
\overline{\rho}^{(n+1)}\widetilde{\rho}^{(1)}
\prod\limits_{i=1}^N
\left(\left(2\lambda^2_i\right)^{3/2}
\exp \left[-\frac{\overline{\bf p}^2_i \lambda_i^2}{2\pi} \right] \right)
\left(\frac{\lambda'_i}{\pi}\right)^3
\delta\left(\widetilde{\bf r}_i^{(n)}-\overline{\bf r}_i^{(1)}\right)
\label{tilde-xi-intl-2}
\end{eqnarray}
\end{widetext}
where $\lambda_i=\sqrt{ \pi \beta /2 m_i }$.
Analytical integration over the delta function simplifies the final
path integral used for further computation of  the real-valued $W_0$ by means of the Monte-Carlo method.

\section{Simulations of QGP}\label{s:model}

\begin{figure}[htb]
\vspace{0cm} \hspace{0.0cm}
\includegraphics[width=7.5cm,clip=true]{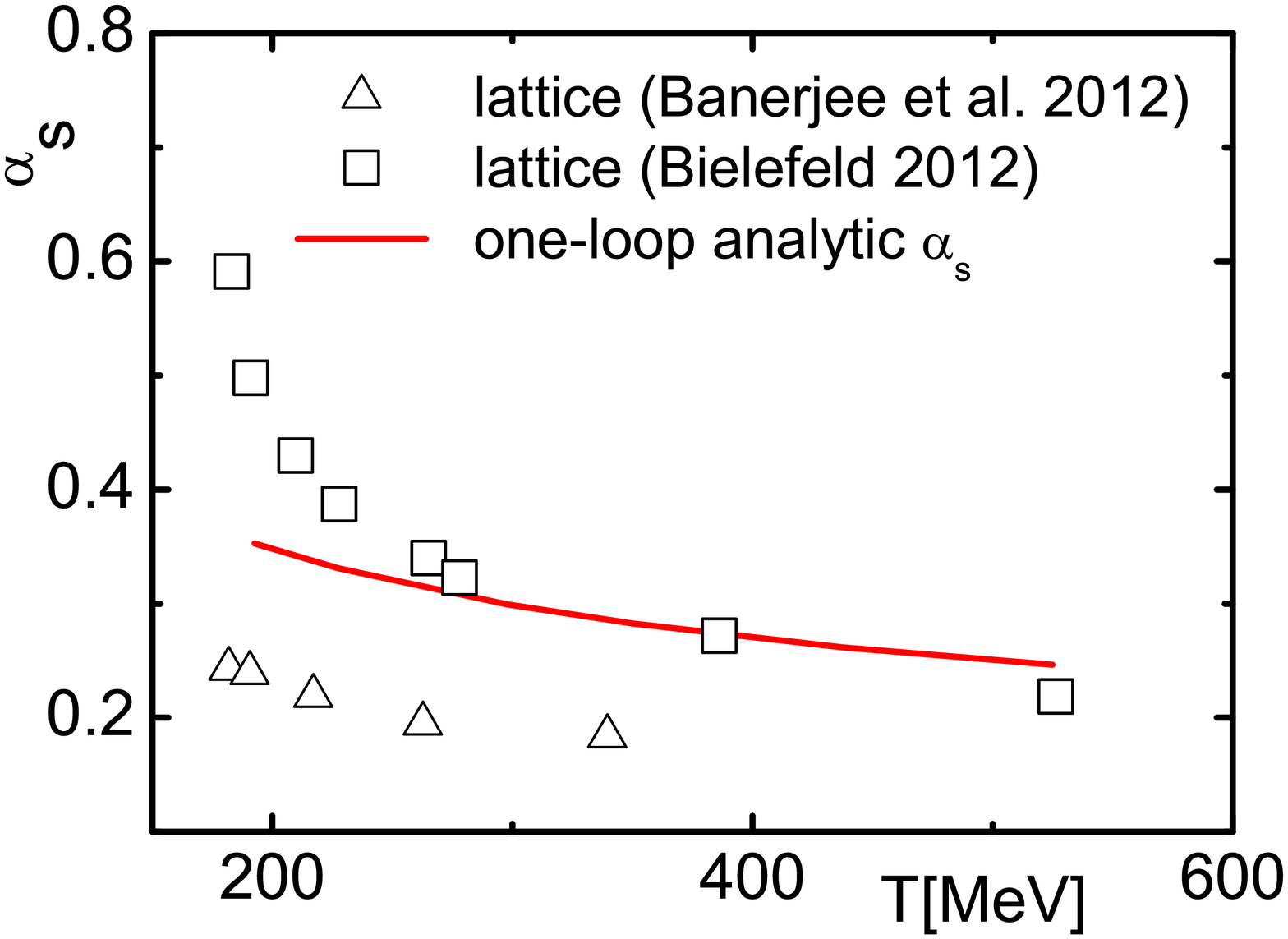}
\includegraphics[width=7.5cm,clip=true]{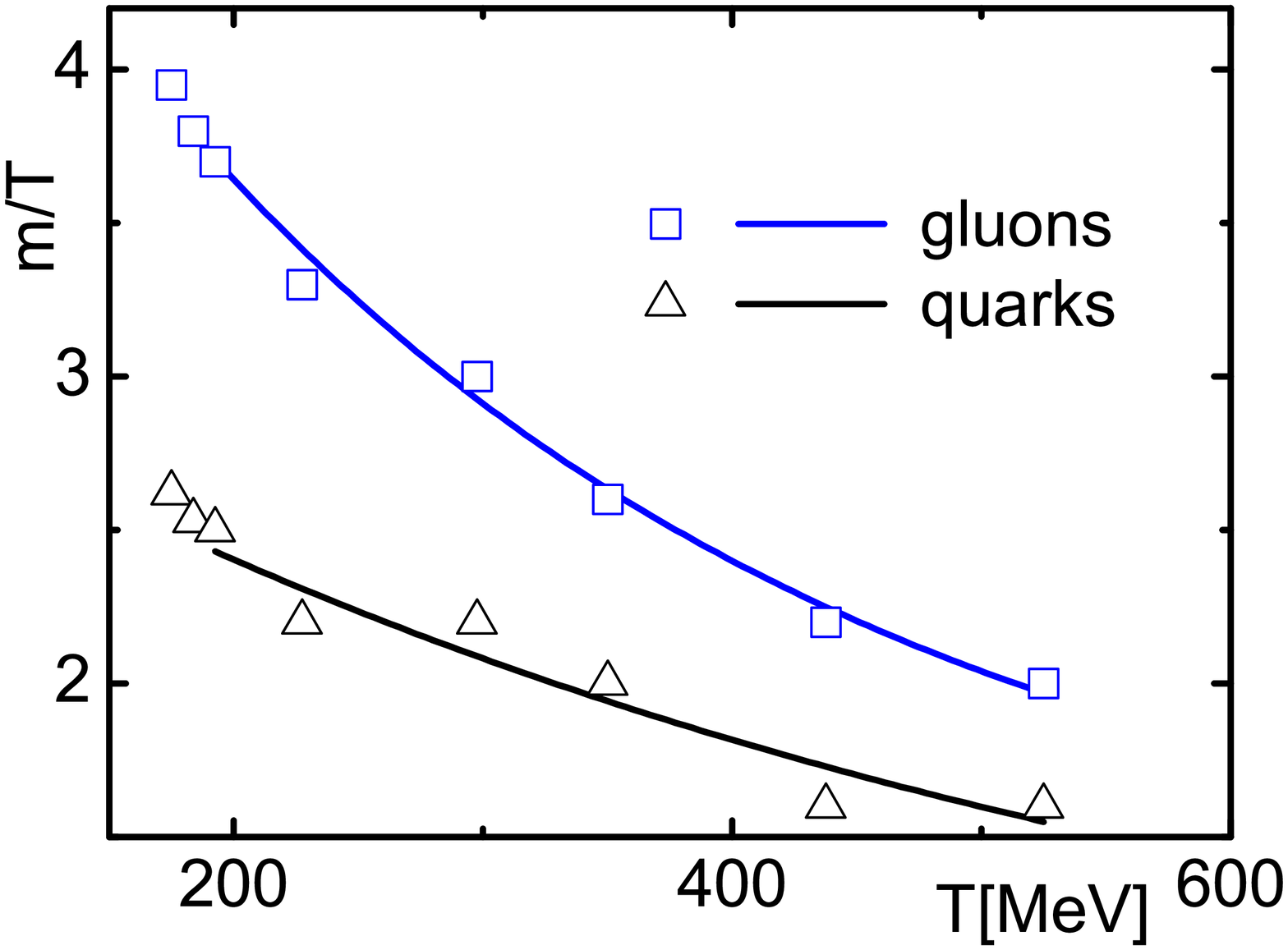}
\caption{(Color online)
Input data for the PIMC simulations.  \\
Panel a: Running coupling constant versus temperature
fitted to experimental data \cite{Shirkov,Prosperi} (solid line).
Points present the coupling deduced from lattice QCD simulations in Refs. \cite{Kaczmarek05} [lattice  (Bielefeld 2012)]
and \cite{Banerjee:2011ra} [lattice  (Banerjee et al. 2012)].
\\
Panel b:
Mass-to-temperature ratio for quark and gluon quasiparticles versus temperature.
Points are values used in simulations.
The solid lines are smooth interpolations  between points.
}
\label{fig:alfrs}
\end {figure}

The developed approach is applied to the QGP at zero baryon density ($\mu_q =0$).
Then assumption on equal quark masses (see point {\bf IV} in  subsect.  \ref{semi:model})
immediately implies equal  fractions of quarks and antiqiarks of different flavors:
$N_u=N_d=N_s=N_q/3=\overline{N}_u=\overline{N}_d=\overline{N}_s=\overline{N}_q/3$.
Ideally the parameters of the model should be deduced from the QCD lattice data. However, presently this task is still quite ambiguous. Therefore, in the present simulations we take only a possible set of parameters.
We use so called ``one-loop analytic coupling'' \cite{Shirkov,Prosperi}
\bea
\alpha_s (Q^2) &=& \frac{4\pi}{11-(2/3)N_f}
\cr
&\times&
 \left[
\frac{1}{\ln (Q^2/\Lambda_{QCD}^2)} + \frac{\Lambda_{QCD}^2}{\Lambda_{QCD}^2-Q^2}
 \right]
\eea
where $Q$ is the momentum transfer,
$\Lambda_{QCD}$ = 206 MeV is the QCD scale and $N_f$ = 3 is the number of flavors.
The analytically generated non-perturbative contribution
${\Lambda_{QCD}^2}/(\Lambda_{QCD}^2-Q^2)$ subtracts
the unphysical Landau pole in a minimal way, yielding a ghost-free behavior which avoids any
adjustable parameter.
This coupling agrees with a great body of experimental data \cite{Shirkov,Prosperi}.
As it is usually done in thermal field models, we substitute
$Q$ by $2\pi T$ to use this coupling in our simulations.
The resulting $\alpha_s (T)$ is displayed in the Panel (a) of Fig.~\ref{fig:alfrs}
and compared with QCD-lattice coupling deduced from a short-distance behavior of the singlet free energy
\cite{Kaczmarek05} and from spectral density of heavy-quark correlator \cite{Banerjee:2011ra}.
As seen, the running coupling deduced from experimental data is close to those
obtained in the lattice QCD. Notice that determination of $\alpha_s (T)$ in lattice QCD simulations
is quite indirect. Therefore, different indirect methods naturally give somewhat different results.

The quasiparticle masses were chosen to reproduce the pressure obtained in lattice QCD calculations \cite{Fodor09,Csikor:2004ik}.
The $T$-dependence of these masses is presented in Fig.~\ref{fig:alfrs}( Panel a).
When choosing masses we kept in mind constraints resulting from
lattice QCD data \cite{Lattice02,Karsch:2009tp,Nakamura03} and QCD-motivated quasiparticle models
\cite{Peshier96,Ivanov05}. While gluon masses used in this paper well comply with those deduced
from both lattice QCD data \cite{Lattice02,Nakamura03} and quasiparticle models, this is not the case
for quark masses. Our quark masses agrees with values required for quasiparticle fits
\cite{Peshier96,Ivanov05} of the lattice  thermodynamic properties
of the QGP:  $m_q/T \simeq 1.5\div 2.5$. At the same time they are appreciably lower than those
in old lattice data  \cite{Lattice02}: $m_q/T \simeq 4$, and higher than $m_q/T \simeq 0.8$ reported
in newer lattice calculations \cite{Karsch:2009tp}.

\subsection{Equilibrium Properties}\label{e:model}

\begin{figure}[htb]
\vspace{0cm} \hspace{0.0cm}
\hspace*{-10mm}
\includegraphics[width=7.5cm,clip=true]{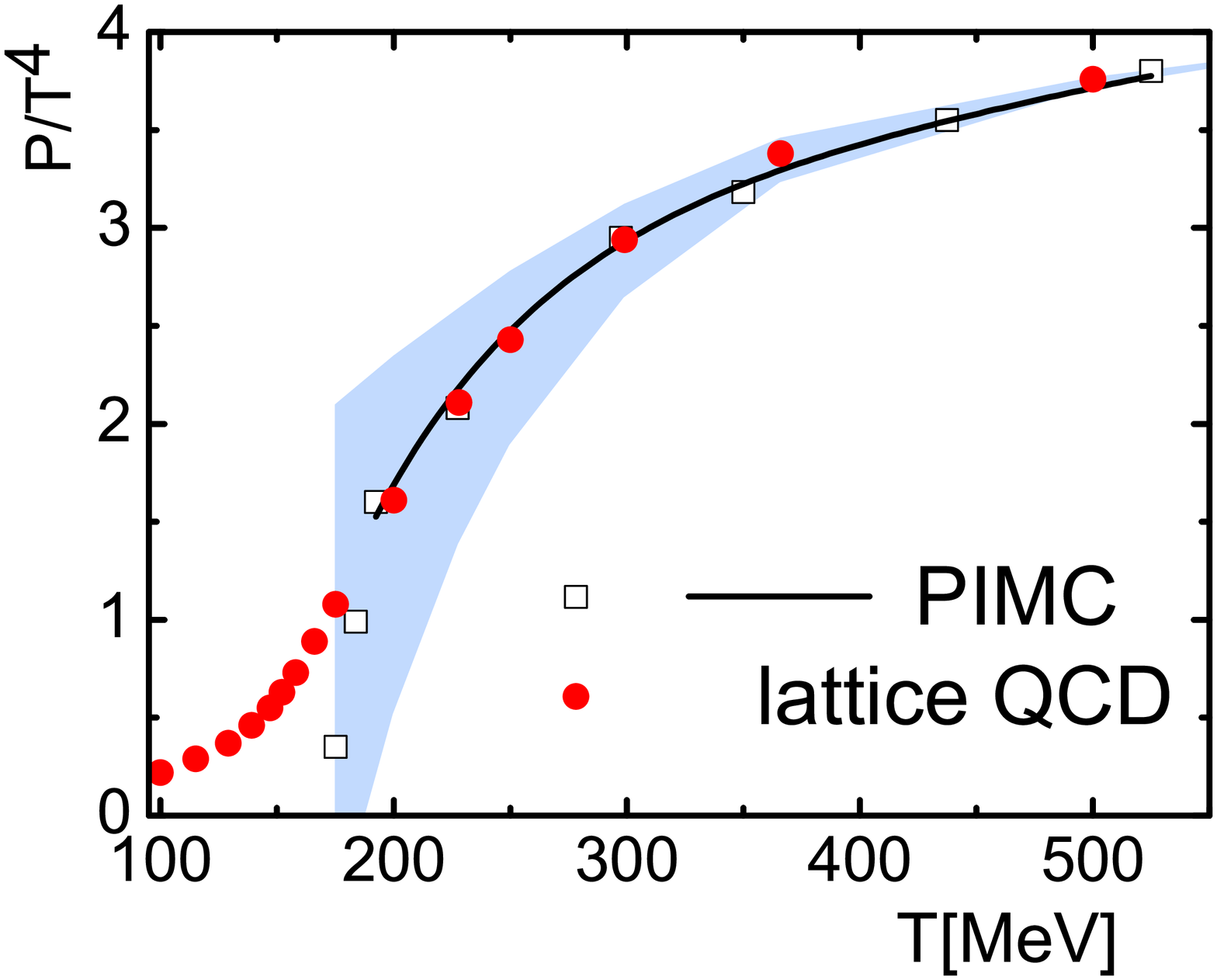}
\hspace*{-10mm}
\includegraphics[width=7.5cm,clip=true]{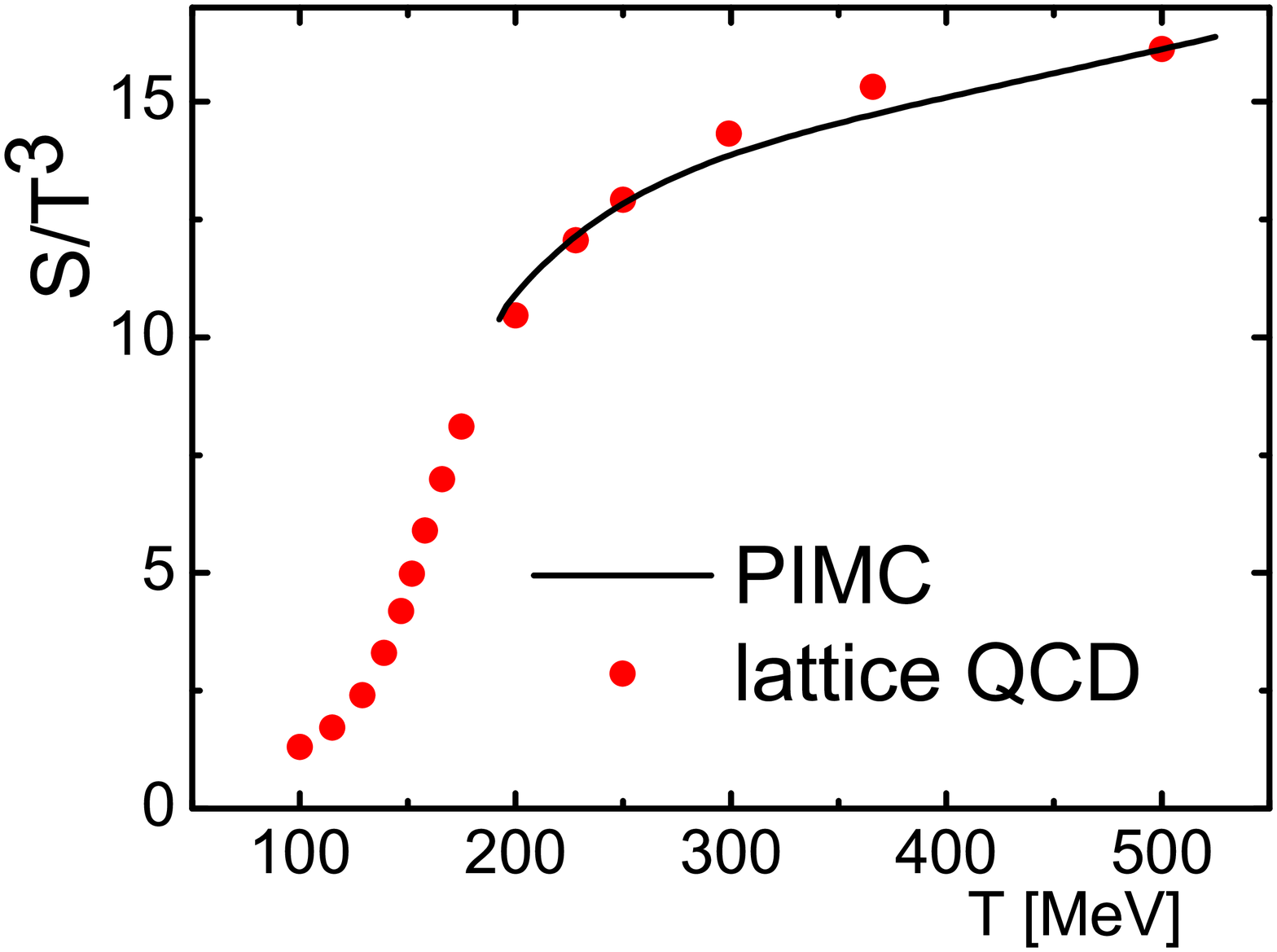}
\hspace*{-10mm}
\includegraphics[width=7.5cm,clip=true]{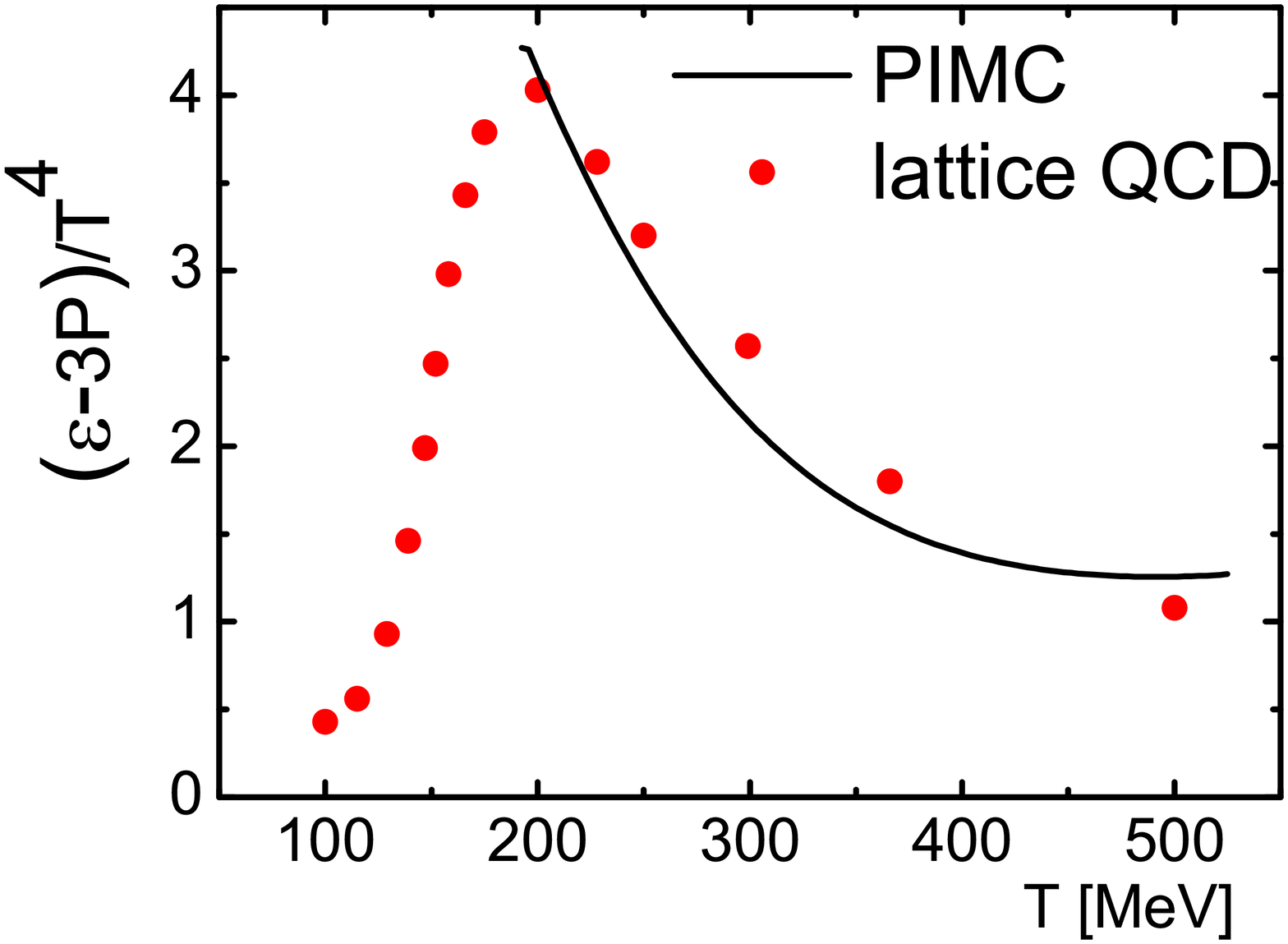}
\caption{(Color online)
Pressure  (Panel a),
entropy (Panel b) and trace anomaly (Panel c)
scaled with corresponding powers of temperature
versus temperature
from PIMC
simulations (open squares). These are compared with
lattice data of  Refs. \cite{Fodor09,Csikor:2004ik} (filled circles).
The solid line in the Panel (a) is a smooth interpolation between the PIMC points.
Entropy and trace anomaly are calculated based on this smooth interpolation.
The shaded aria in the Panel (a)  indicates uncertainties of the PIMC calculations.
}
\label{fig:EOS}
\end {figure}

Figure \ref{fig:EOS} (Panel a) demonstrates the quality of reproduction of the equation of state (EOS),
i.e. the pressure versus temperature, achieved with the above discussed input data.
The reference EoS (filled points in the Panel (a) of Fig.~\ref{fig:EOS}) is taken from
QCD lattice simulations of the QGP \cite{Fodor09}. The quality of the reproduction obviously
depends on the degree of accurate tuning of the input data, i.e. the quasiparticle masses.
However, the PIMC scheme itself also produces certain errors. If there are
metastable states of the system, convergence of calculations
 becomes poor because of
 jumps between stable and metastable states. This is a typical
situation when the
system approaches to a point (or a range) of a phase transition. Precisely this happens at the
lower end of considered temperature range.
The shaded aria in the Panel (a) of Fig.~\ref{fig:EOS} indicates these
uncertainties of the PIMC calculations.
Figure \ref{fig:EOS} also presents the entropy $S/T^3$ and
trace anomaly $(\varepsilon-3P)/T^4$  of the QGP. These quantities are calculated accordingly to Eqs.
(\ref{p_gen})-(\ref{e_gen}). In order to avoid the numeric noise, the derivative of
a smooth interpolation between the
PIMC points (solid line in the Panel (a) of Fig. \ref{fig:EOS})  was taken.
Though agreement with the lattice data looks worse for the entropy and
trace anomaly, in fact it is the same as that for pressure.  Differentiation operations
in Eqs. (\ref{s_gen}) and (\ref{e_gen}) make differences between PIMC results and
lattice data more pronounced.

\begin{figure}[tb]
\vspace{0cm} \hspace{0.0cm}
\includegraphics[width=7.9cm,clip=true]{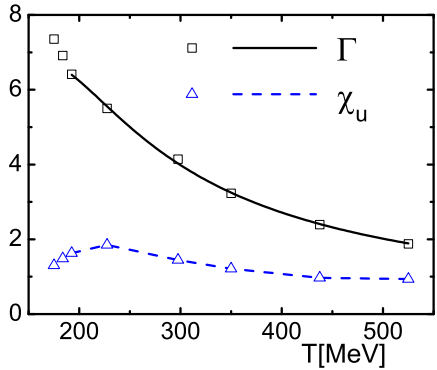}
\caption{(Color online)
Quark degeneracy parameter $\chi_u$
and the plasma coupling parameter $\Gamma$ [see Eq. (\ref{Gamma})]  versus temperature.
Lines are smooth interpolations  between points.
}
\label{fig:EOS1}
\end {figure}

Having calibrated the model by reproducing the EoS, we can proceed to predictions.
First, let us consider internal properties of the system.
To characterize physical conditions and interplay of interaction
and degeneracy  in  Fig.~\ref{fig:EOS1} a degeneracy parameter
$\chi_u$ for 'up' quarks and a
plasma coupling parameter $\Gamma$ are presented:
\begin{eqnarray}
\label{Gamma}
\chi_u=n_u\lambda^3_u,   \quad \quad
\Gamma = \frac{ \overline{q}_2 g^2}{4\pi r_s T},
\label{chi-gam}
\end{eqnarray}
where the thermal wave length $\lambda_u$ was defined in the
previous sectios (see text after Eq. (\ref{Lambda})),
$n_u$ is density of $u$ quarks,
$r_s^3=3/(4\pi n)$ is Wigner-Seitz radius, $n$ is the density of all quasiparticles
(quarks, antiquarks and gluons), and
$\overline{q}_2$ is the quadratic Casimir value
averaged over quarks, antiquarks and gluons,
$\overline{q}_2=N_c^2-1$ is a good estimate for this quantity.
The plasma coupling parameter is a measure of ratio of the average potential
to the average kinetic energy, and the degeneracy parameter $\chi_u$
indicates wheather a system is classical ($\chi_u\ll 1$) or quantum ($\chi_u\gsim 1$).
It turns out that $\Gamma$ and $\chi_u$ are of order
unity which indicates  that the QGP is a strongly coupled
{\em quantum ($\chi_u\gsim 1$) liquid ($\Gamma \sim 1$)}
rather than a gas.

\begin{figure}[phtb]
\vspace{0cm} \hspace{0.0cm}
\includegraphics[width=7.4cm,clip=true]{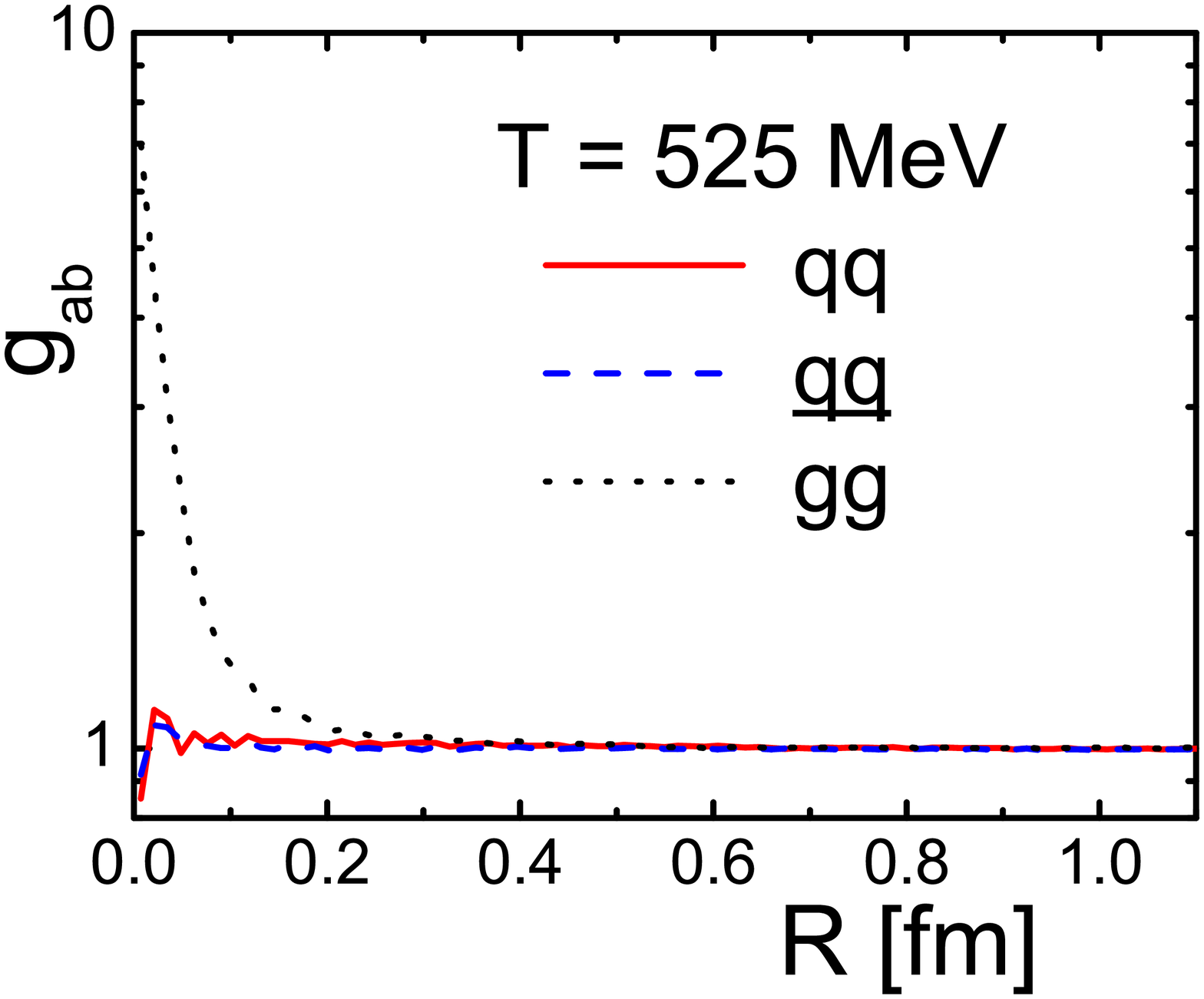}
\includegraphics[width=7.4cm,clip=true]{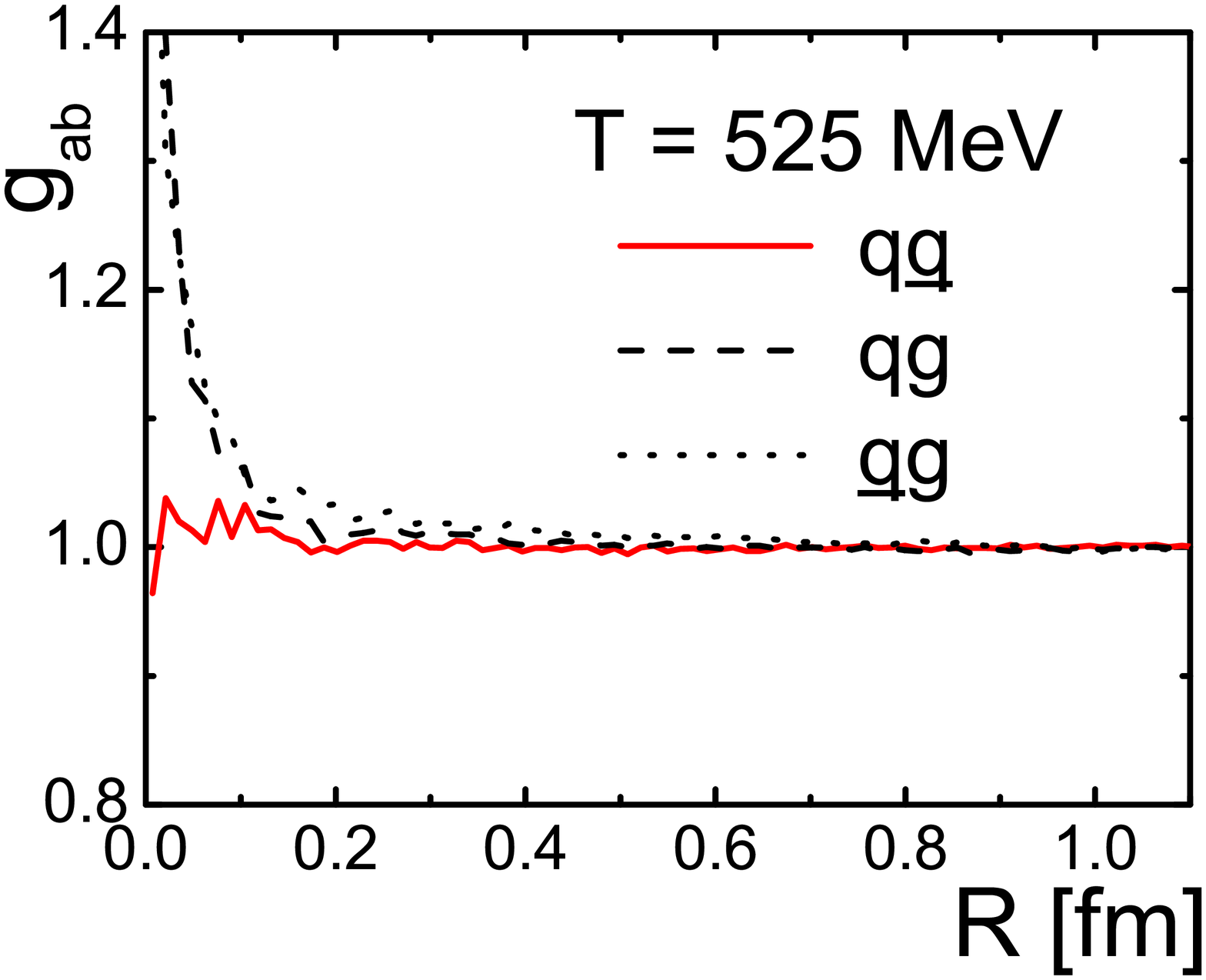}
\includegraphics[width=7.4cm,clip=true]{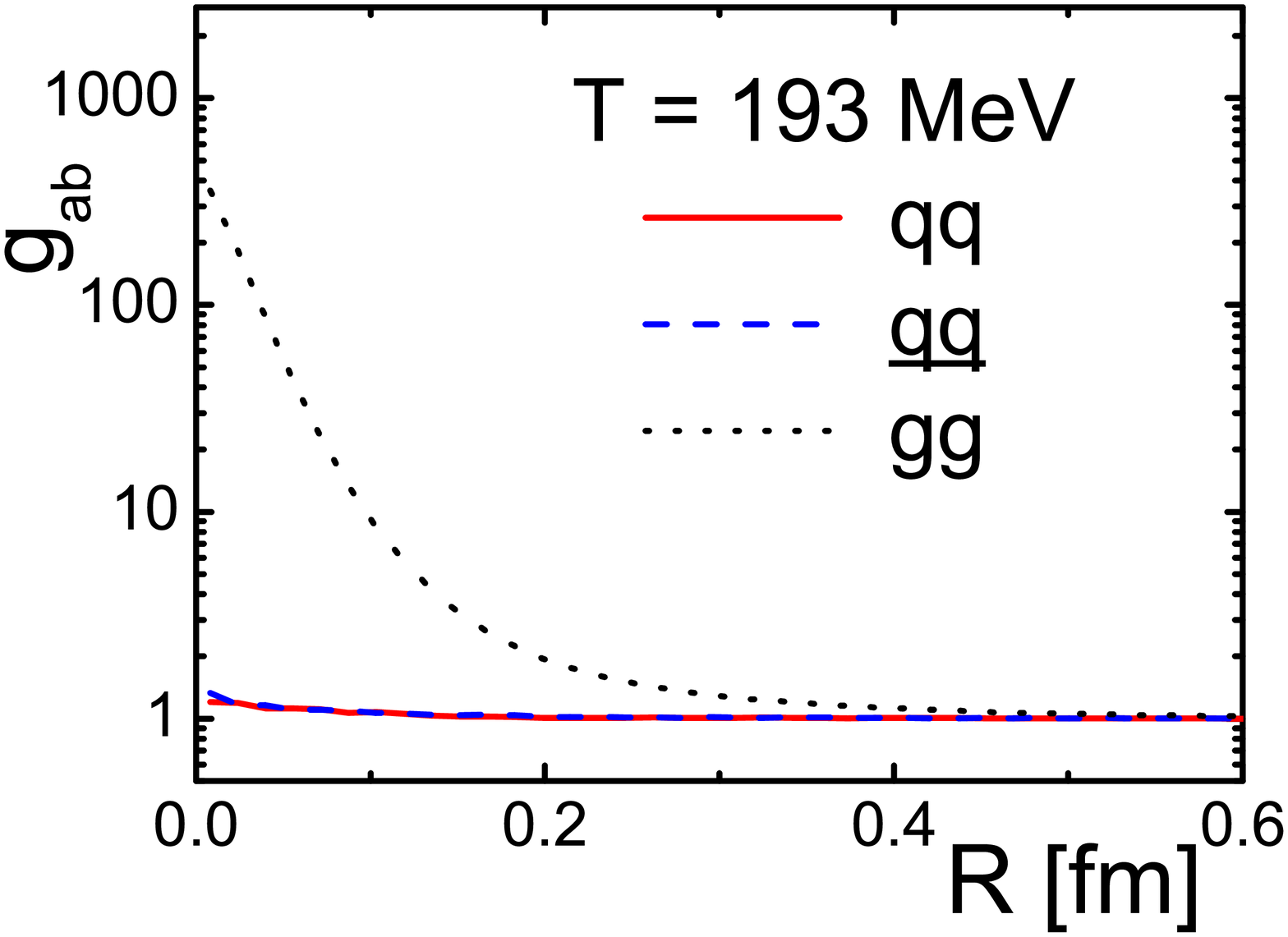}
\includegraphics[width=7.4cm,clip=true]{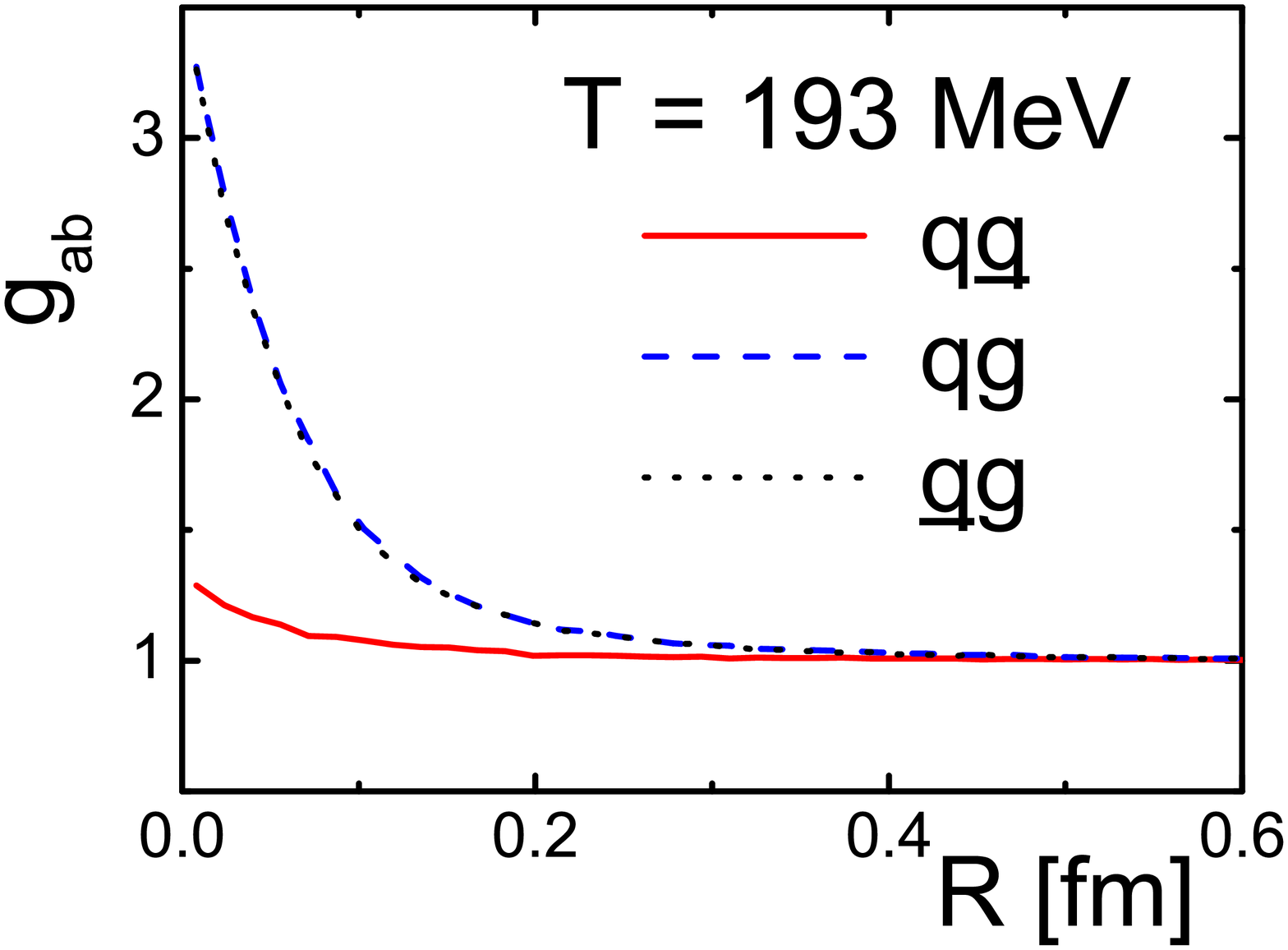}
\caption{(Color online)
Pair correlation functions
of identical (Panels a, c) and different (Panels b, d) quasiparticles  at
temperatures $T=525 \text{ MeV}$ (Panels a, b) and $T=193 \text{ MeV}$ (Panels c, d ).
}
\label{fig:PDFC}
\end{figure}

To clarify interplay of interaction and degeneracy
let us consider spatial arrangement of the quasiparticles in
the QGP by studying a pair distribution function (PDF) $g_{ab}(R)$ defined as
\begin{widetext}
\begin{eqnarray}\label{g-def}
g_{ab}(|{\bf R}_1-{\bf R}_2|) &=&
\left(\frac{V}{N}\right)^2 
\sum_{\sigma}
\sum_{i,j,i\neq j}
\delta_{a_i,a}\, \delta_{a_j,b}\frac{1}{Z}
\int
dr\; d\mu Q \,\delta({\bf R}_1-{\bf r}_i)\,\delta({\bf R}_2-{\bf r}_j)\;\rho(r,Q, \sigma; 
\{N\}; \beta),
\end{eqnarray}
\end{widetext}
where $a_i$ and $a_j$ are types of the particles ($=q,\overline{q}$ or $g$).
The PDF gives a probability density to find a pair of particles of types
$a$ and $b$ at a
certain distance $R=|{\bf R}_1-{\bf R}_2|$ from each other.
The PDF depends only on the difference of coordinates because of
the translational invariance of the system.
In a non-interacting classical system,
$g_{ab}(R)\equiv 1$, whereas interactions and  quantum statistics result in
a redistribution of the particles.
At temperatures $T=525 \text{ MeV}$ and $T=193 \text{ MeV}$  the
PDF averaged over the quasiparticle spin, colors and flavors are shown in 
Fig.~\ref{fig:PDFC}.
\begin{figure}[htb]
\vspace{0cm} \hspace{0.0cm}
\includegraphics[width=7.4cm,clip=true]{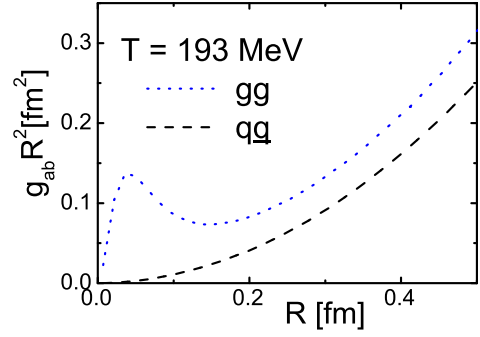}
\caption{(Color online)
Gluon-gluon and quark-antiquark pair correlation functions multiplied by $R^2$
at $T=193 \text{ MeV}$.
}
\label{fig:PDFC_r2}
\end{figure}

At distances $R \gsim$ 0.2 or 0.3 fm, depending on the temperature,
all PDF's are practically equal
 to unity (Fig. \ref{fig:PDFC}) like in ideal gas due to the screening of the
  color Coulomb interaction.
A drastic difference between $qq$ and $gg$
PDF's  (the $\overline{q}\overline{q}$ PDF is identical to the $qq$ one)
occurs at short distances.
Here the gluon-gluon and gluon-quark PDF's
increase monotonically when the distance goes to zero while the $qq$ and $\overline{q}\overline{q}$ ones
remain uncorrelated.
One of the physical reasons of the PDF difference is spatial quantum uncertainty
and different properties of Bose and Fermi statistics of gluon and (anti)quark
quasiparticles. Uncertainity in particle
localization is defined by the ratio $T/m$. Localization is better for
heavier gluon quasipartcles.
Fermi statistics results in effective quark-quark and
aniquark-antiquark repulsion, while Bose one results in effective qluon-qluon
attraction.
Oscillations of the PDF at very small distances
$R\lsim 0.1 $ fm are related to Monte-Carlo statistical error,
as probability of  quasiparticles being at short distances quickly decreases.

However, the $qq$ and $\overline{q}\overline{q}$ pair correlation functions reveal practical absence
of fermi repulsion.
This happens because another physical reason comes into play.
Strong interaction
between quasiparticles reduces the influence of the degeneracy in
the region of $\chi_u \sim 1$.
This interaction is dominated by attraction at short distances.
Indeed,
the QGP lowers its total energy by minimizing the color
Coulomb interaction energy via a spontaneous ``anti-ferromagnetic''-like ordering of color vectors,
i.e.  the color vectors of nearest neighbor quasiparticles become anti-parallel.
Similar  absence of fermi repulsion was observed in hydrogen plasma
 at $\chi \sim 1$ \cite{zamalin}.
This short-distance attraction is stronger for gluon-gluon and gluon-(anti)quark pairs
than for (anti)quark-(anti)quark ones because of the corresponding
difference in values of quadratic Casimir invariants $\breve{q}_2$ (see Appendix I),
which determine the maximal values of the effective color charge products
$\left|Q_i\cdot Q_j\right|$ in color Kelbg (Coulomb) potentials:
For gluon-gluon pairs $\left|Q_g\cdot Q_g\right|_{max}=$ 24,
for gluon-(anti)quark pairs
$\left|Q_g\cdot Q_q\right|_{max}=\left|Q_g\cdot Q_{\overline{q}}\right|_{max}\approx$ 10,
and for (anti)quark-(anti)quark pairs
$\left|Q_q\cdot Q_q\right|_{max}=\left|Q_{\overline{q}}\cdot Q_{\overline{q}}\right|_{max}=\left|Q_q\cdot Q_{\overline{q}}\right|_{max}=$ 4.
Stronger $gg$ attraction additionally enhances correlation of the gluon-gluon pairs
at short distances. At the same time the short-distance attraction is the only
reason of the gluon-(anti)quark short-distance correlation.

The short-distance correlation implies formation
 of the gluon-gluon and gluon-(anti)quark clusters, which are uniformly distributed in space.
In case of the gluon-gluon clusters we can even speak about $gg$ bound states, i.e. glueballs,
as it is seen from Fig. \ref{fig:PDFC_r2}.
The product $R^2 g_{ab}(R)$ is proportional (up to constant factor) to
a probability to find
a pair of quasiparticles at a distance $R$ from each other.
As known from consideration of hydrogen plasma
\cite{zamalin}, a maximum in
$R^2\,g_{gg}(R)$ signals population of a bound state.
For comparison, the quark-antiquark correlation function, i.e.  $R^2\,g_{q\overline{q}}(R)$,
is also presented in Fig. \ref{fig:PDFC_r2}. It demonstrates that there are no bound
meson-like states. We can only speak about weak meson-like clustering at
lower temperatures, see short-distance $q\overline{q}$ correlation at T=193 MeV in Fig. \ref{fig:PDFC}.
Possible existence of medium-modified meson-like bound states was actively
discussed some time ago, e.g.,  in  Ref. \cite{Yukalov97} and later in  Refs. \cite{Shuryak03,Brown05}
based on results from lattice QCD calculations
of spectral functions \cite{Asakawa01,Karsch03}.
Our result supports conclusion of Ref.
\cite{Koch05} on the absence of $q\overline{q}$ bound states above the temperature of the phase transition.
This finding is in contrast to our previous results on SU(2)
 group \cite{Filinov:2012pimc,Filinov:2010pimc,Filinov:2012zz,Filinov:2009pimc}.
There well pronounced bound $q\overline{q}$  states were found just above the critical temperature,
which however quickly dissolved with the temperature rise.
This happens because the SU(3) plasma turns out to be essentially denser than the SU(2) one,
which is a consequence of a stronger effective attraction between constituents.
As a result, possible bound states in the SU(3) plasma just melt.
To verify the relevance of all above discussed trends, a more
refined color-, flavor-, spin-resolving analysis of the PDF's is necessary.
This work is presently in progress.

{
\subsection{Monte Carlo Simulations}\label{MC}
}

{
Details of our path integral Monte-Carlo simulations have been discussed elsewhere
in a number of papers and review articles, see,
e.g. Refs. \cite{zamalin,rinton} and references therein.
For simulation of the thermodynamic properties of QGP we use the standard Metropolis
algorithm. We use a cubic simulation box with periodic boundary conditions.
The main idea of the simulations consists in constructing a Markov
chain of different quasiparticle states in the configuration space including the color.
The computational procedure consists of two stages.}

{
At the first stage a dominant, i.e. maximal, $\{N\}$-term in the sum of Eq. (\ref{Gq-def})
is determined by calculations in grand canonical ensemble.
This term is indeed dominant in the thermodynamic limit of the box volume $\to \infty$.
In grand canonical ensemble  the quasiparticle numbers in simulation box are varied, 
i.e. the consecutive states of the Markov chain can differ from each other by numbers of
quarks, antiquarks and gluons. Transitions
between these states are the first type of
markovian elementary steps. In the second type of elementary steps
coordinates of a single bead of a randomly chosen quasiparticle are changed.
The color variables are changed accordingly to the SU(3) group Haar measure
in the third type of markovian elementary steps.
We generate the Markov chain until a full convergence of calculated values is achieved.
Thus, we determine the average numbers
of quarks, antiquarks and gluons in the box at fixed temperature. Here only 
 densities of each species, i.e.  the ratios of the 
these average numbers to the box volume, have the physical meaning. 
Usually, after several millions of elementary steps
the average numbers of these quasiparticles become stable and the average number of
quarks practically equals that of antiquarks. This equality is
considered as an inherent test of consistency of the calculations at zero baryon
chemical potential.}

{
 At the second stage we fix the number of quarks, antiquarks and gluons to be equal
to the obtained average values and
perform calculations in the canonical ensemble.
Here we use only the second and third types of the elementary steps described above.
We calculate the pressure defined in Eq. (\ref{eos})
 An important difference from the case of the
electrodynamic plasmas consists in using the relativistic measure in path integrals.
This measure is associated with  relativistic kinetic energy operator instead of the
conventional Gaussian one arising from the non-relativistic operator of kinetic
energy in Feynman-Wiener path integrals.
After several millions of elementary markovian steps
the result for the pressure becomes stable.
}

{
Errors of Monte-Carlo calculations of thermodynamic quantities related
to the finite particle number ($N$) in the system with periodic boundary
conditions are of the order of $1/N$ \cite{zamalin}.
However, too large number of particles presented by a large number of beads
requires too large computer resources. 
In practical calculations we try to keep the total number of particles  not exceeding
$N=N_q+N_{ \overline{q}}+N_g=126$ and adjust the above determined proper densities  
of species by varying the total volume $V$ of the box. The number of beads $n=20$ for
each particle is used.
Our choice of particle and bead numbers
is a compromise between acceptable accuracy 
and available computer resources. It was checked that variation of the
number of beads from $15$ up to $50$ practically does not change results.}

{
As it follows from Fig. \ref{fig:EOS1}, the degeneracy is moderate in our case,
i.e. the degeneracy parameter is of order of several units.
Therefore, the well known sign problem in Monte-Carlo simulations
of Fermi particles is not very severe here. In our calculations we reduce the sign problem
from the level of sign interference of the permutations to the level
of sign interference of determinants. For this purpose we include the modulus of determinants
of Eq. (\ref{Grho_s}) in the probability of the markovian elementary steps, while
the sign of the determinants is attributed to the weight function at calculations
of the average quantities. Thus, each markovian step is equivalent to the $N!$
markovian steps in the sum over permutations. This method was tested at the example 
of the ideal Fermi gas \cite{zamalin}. It was found that the method results in 
agreement with the exact solution up to 
values of $\sim$ 10-15 of the degeneracy parameter (\ref{chi-gam}), if the particle wave 
length is smaller than the size of Monte Carlo box. 
As shown in calculations of Ref. 
\cite{zamalin,fusion,Fehske},
this method works well enough for hydrogen and electron-hole plasmas.
We anticipate that this approach will be efficient at least at moderate values of 
the baryon chemical potential, i.e. up to $\mu_q \sim T$. 
\\}

\subsection{Transport coefficients}\label{k:model}

An important aspect of the strongly coupled QGP is its transport properties which strongly differ from those
we would expect for weakly coupled plasmas.  We use the developed approach based on Wigner formulation of
quantum mechanics to calculate the QGP transport properties at strong coupling. In particular, we
calculate the QGP self-diffusion constant and shear viscosity, as these quantities can be compared to
respective values deduced from analysis of
experimental data heavy-ion collisions
and also predictions of the lattice QCD computations.
More precisely, summary of shear viscosity deduced from analysis of experimental elliptic flow
is presented in Ref. \cite{Bass:2011zz}, in Ref.
\cite{Khvorostukhin:2010aj} an extensive review of theoretical works on viscousity is done,
while the heavy-quark diffusion constant is available from the experiment analysis
\cite{Gossiaux:2012ea,He:2012df}
and QCD lattice computations \cite{h-q-diff-latt-11,Banerjee:2011ra}.
We anticipate that the self-diffusion and heavy-quark diffusion constants are compareable by the order of magnitude.

\begin{figure}[htb]
\vspace{0cm} \hspace{0.0cm}
\includegraphics[width=7.5cm,clip=true]{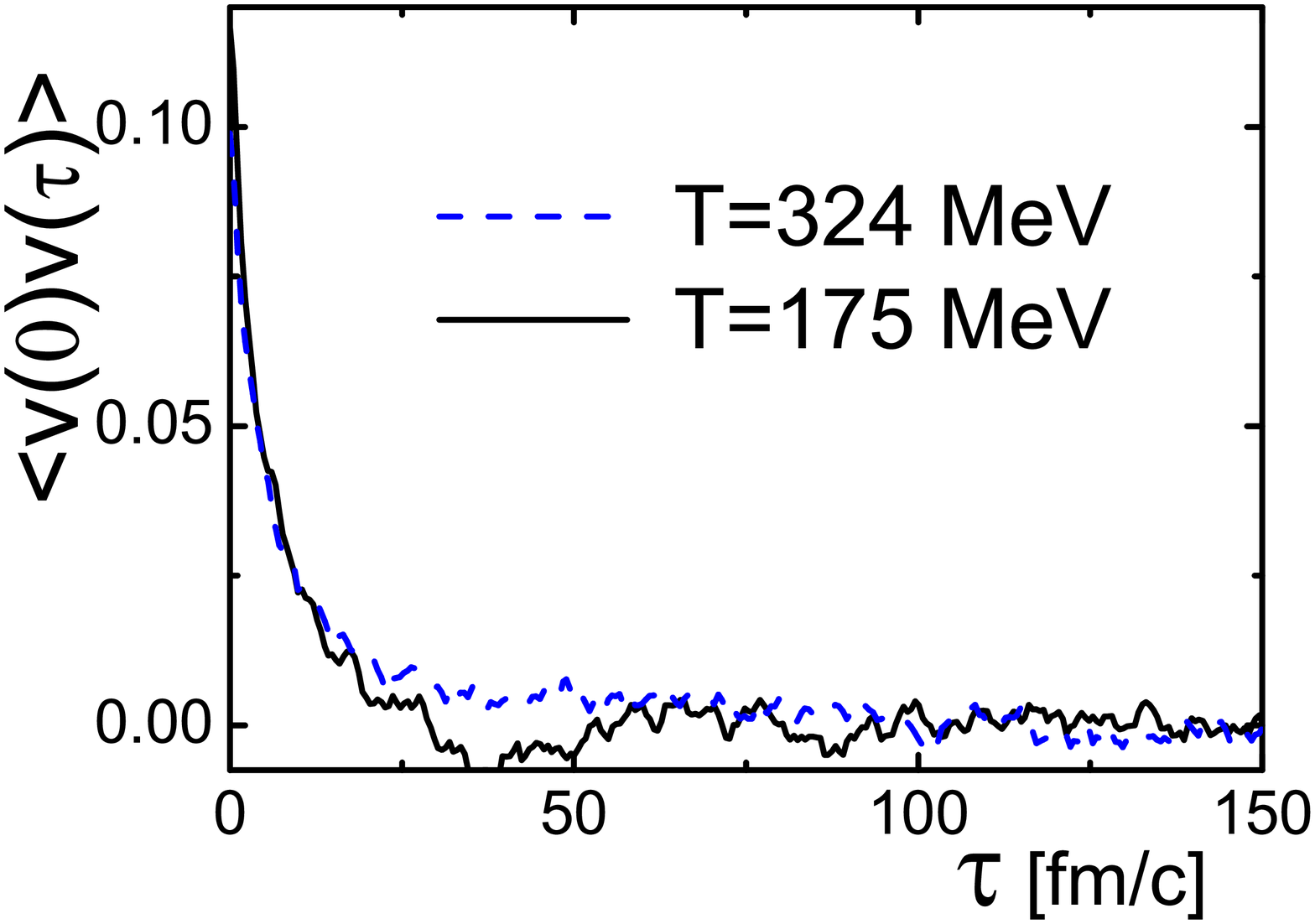}
\includegraphics[width=7.5cm,clip=true]{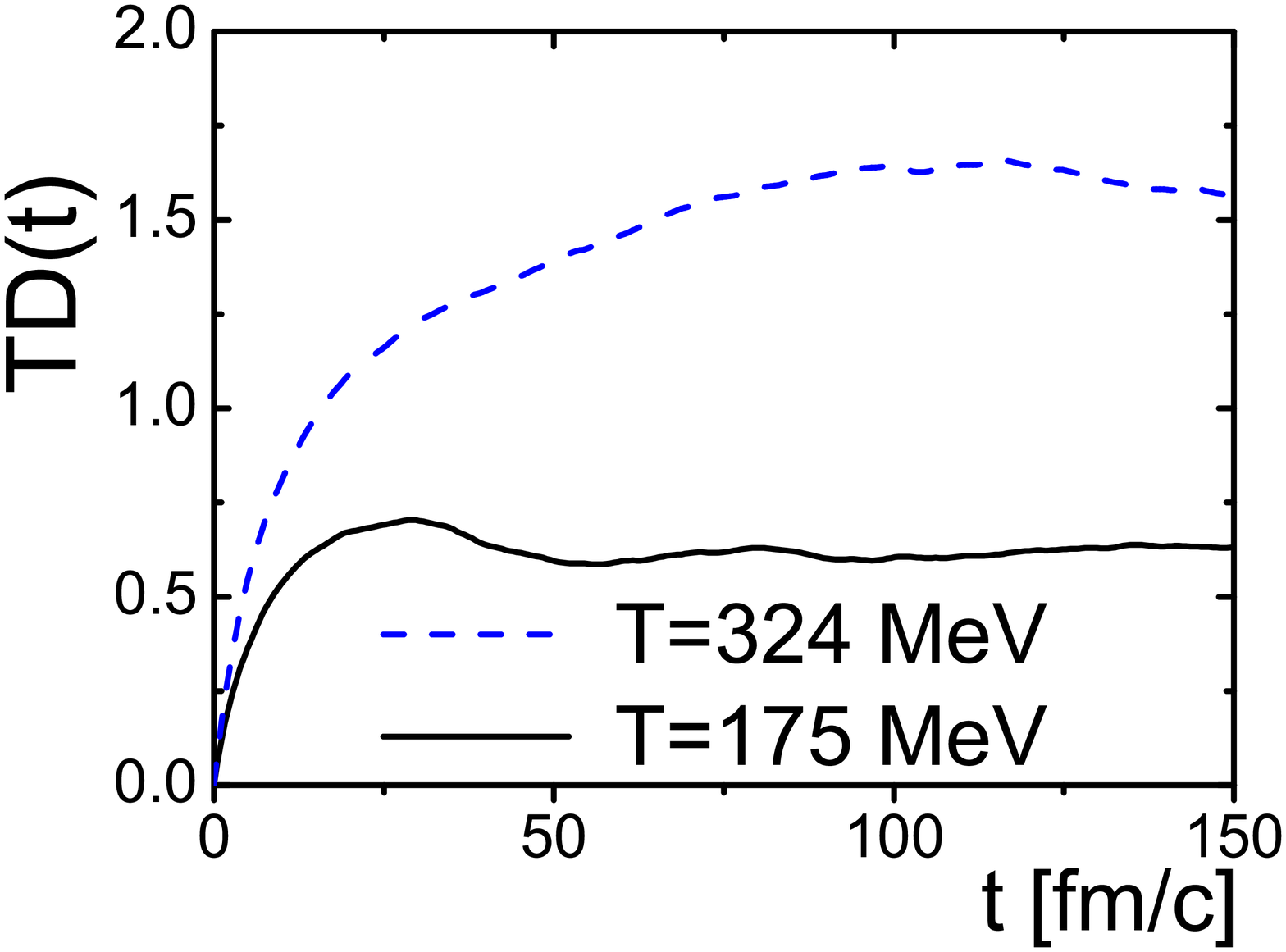}
\caption{(Color online)
Velocity autocorrelation function (Panel a)
and temperature-scaled self-diffusion function
$T\,D(t)$ (Panel b) versus time for two temperatures.
}
\label{fig:VV}
\end {figure}

A natural way to obtain these transport coefficients
is use of the quantum Green-Kubo relations. These relations give the transport coefficients in terms of integrals of equilibrium time-depended correlation functions.
According to the Eq. (\ref{CFA}) a self-diffusion  constant $D$ is the integral of the velocity autocorrelation function 
\bea
D&=&\lim_{t\rightarrow\infty}D(t), 
\cr
D(t)&=& \frac{1}{3}\int_{0}^t d\tau \langle v(0)\cdot v(\tau)\rangle
\cr
\langle v(0) \cdot v(\tau)\rangle&=& (2\pi)^{-6N}
\int d \overline{pq\mu Q}\,  d \widetilde{pq \mu Q} 
\cr
&\times&
W\!\left(\overline{pqQ};\widetilde{pqQ};\tau;\beta \right) \,
v\left(\overline{p}(\tau )\right) \cdot v\left(\widetilde{p}(\tau )\right) ,
\cr
&&
\label{difconst}
\eea
where the product of 3-velocities is
\bea
&&
v\left(\overline{p}(\tau )\right) \cdot v\left(\widetilde{p}(\tau )\right)
\cr
&=&\frac{1}{N}\sum_{i=1}^N
\frac{\overline{\bf p}_i(\tau)  \cdot \widetilde{\bf p}_i(\tau ) }
{\sqrt{\overline{\bf p}^2_i(\tau ) +m_i^2} \sqrt{\widetilde{\bf p}^2_i(\tau) +m_i^2}}.
\label{autovv}
\eea
{
where the spectral density $W\!\left(\overline{pqQ};\widetilde{pqQ};\tau;\beta \right)$ is given by
Eq. (\ref{1st-w}), while trajectories in positive (bared) and inverse (tilded) time
directions are defined by Eqs. (\ref{HW-dir}) and (\ref{HW-inv}), respectively.
Figure~\ref{fig:VV} shows examples of the velocity-velocity autocorrelation and 
its antiderivative functions. The self-diffusion constant is a limiting value of 
the related antiderivative function at $t\rightarrow\infty$.}

{
Calculations of autocorrelation functions are performed in canonical ensemble and include 
combination of the Monte-Carlo sampling of initial conditions $\overline{pqQ}_0$ and 
$\widetilde{pqQ}_0$ for trajectories and solving the system of dynamic equations (\ref{HW-dir}) 
and (\ref{HW-inv}). 
The initial conditions $\overline{pqQ}_0$ and $\widetilde{pqQ}_0$ for the trajectories 
are sampled 
by Monte-Carlo method accordingly to the probability
 $W_0\!\left(\overline{pqQ}_0;\widetilde{pqQ}_0;\beta \right)$. 
The autocorrelator (\ref{autovv}) as a function of time is calculated along the 
trajectories (\ref{HW-dir}) and (\ref{HW-inv}),
which themselves are computed  by means of a numerical scheme for solution of 
a system of ordinary differential equations of the first order. We use the explicit 
numerical scheme with automatically adapted time step. To check correctness of the calculations
we control values of three integrals of motion: the energy, and
quadratic and cubic Casimirs. Their variations
in our calculations amount to less than 1-2\%.
Usually several  thousands of generated trajectories are required
for convergence of the antiderivative of the 
autocorrelation function up to accuracy of 5-10\%.
The convergence is fast enough because the autocorrelation function includes
averaging-out (i.e. summation) over all quasiparticles.  }

\begin{figure}[htb]
\vspace{0cm} \hspace{0.0cm}
\includegraphics[width=7.5cm,clip=true]{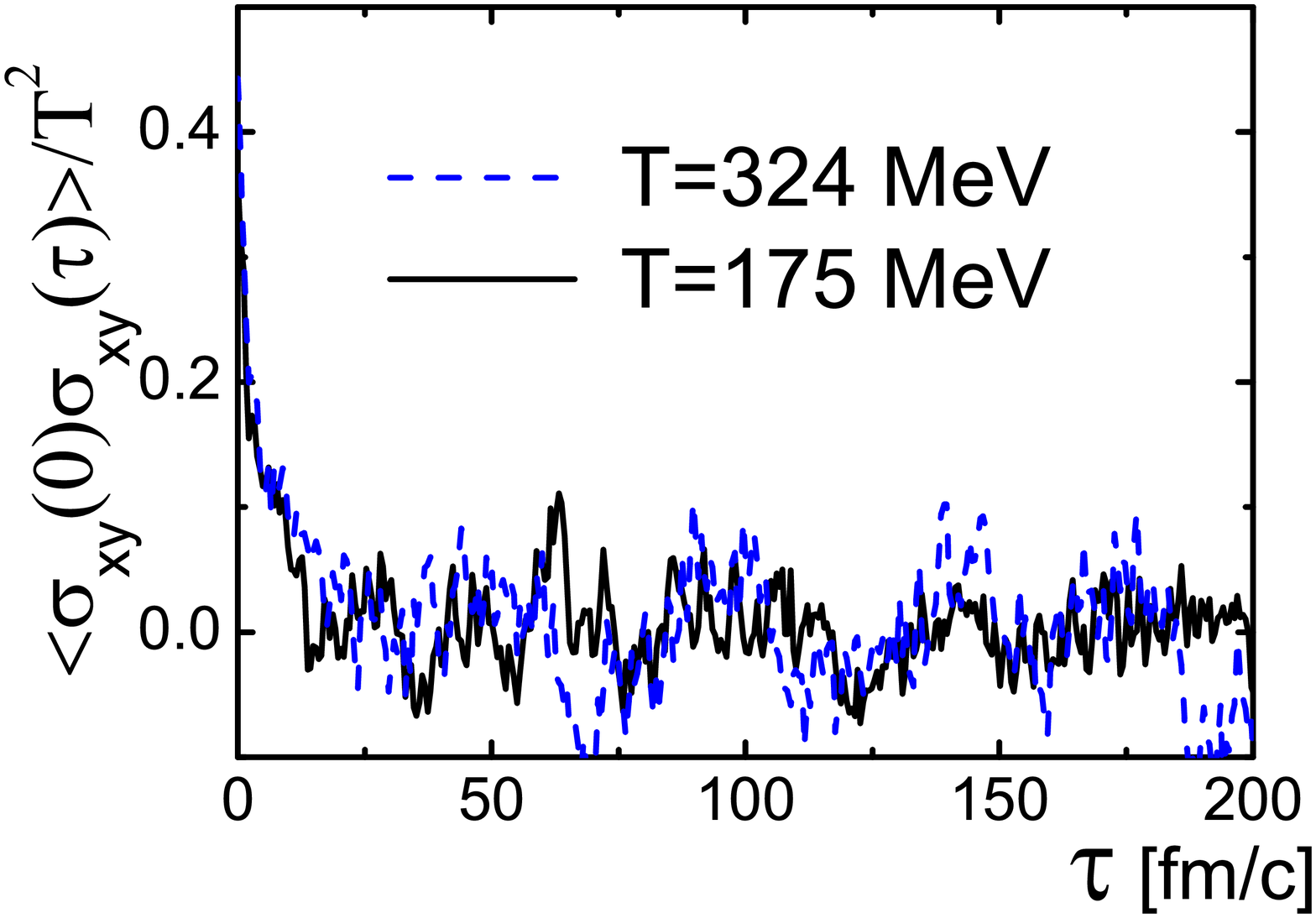}
\includegraphics[width=7.5cm,clip=true]{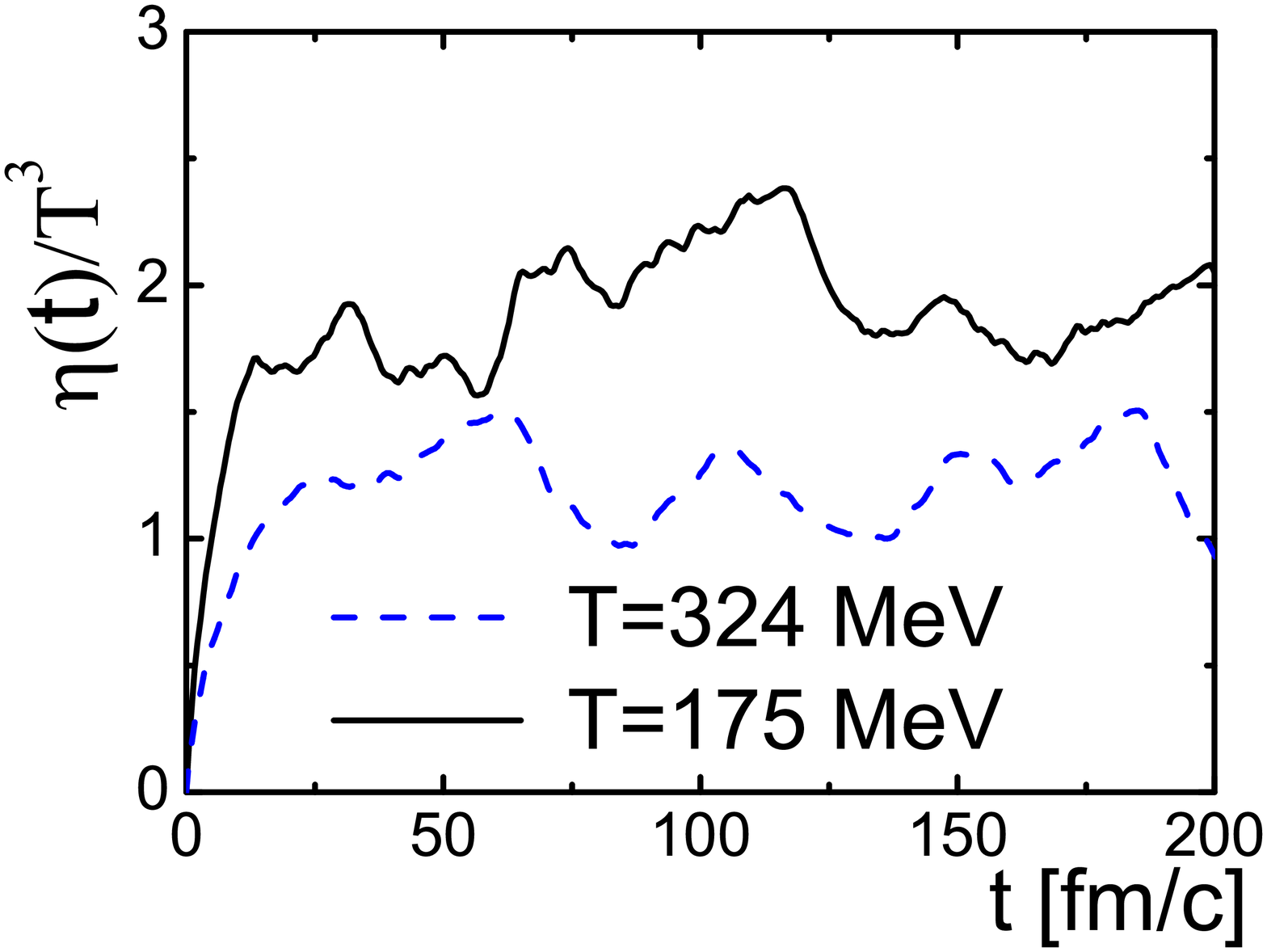}
\caption{(Color online)
Temperature-scaled stress-energy-tensor autocorrelation function (Panel a)
and shear-viscosity  function $\eta(t)/T^3$ (Panel b) versus time
for two  temperatures.
}
\label{fig:StrVis}
\end {figure}

Analogously, the Green-Kubo relation for the shear viscosity is the integral of the autocorrelation function of the stress-energy tensor 
\bea
\eta&=&\lim_{t\rightarrow\infty}\eta(t), 
\cr
\eta(t)&=& \frac{1}{VT}\int_{0}^t d\tau \langle \sigma_{xy}(0) \, \sigma_{xy}(\tau)\rangle
\cr
\langle \sigma_{xy}(0) \,\sigma_{xy}(\tau)\rangle&=& (2\pi)^{-6N}
\int d \overline{pq\mu Q}\,  d \widetilde{pq \mu Q}\,
\cr
&\times&
W\left(\overline{pqQ};\widetilde{pqQ};\tau;\beta \right) \,
\cr
&\times&
\sigma_{xy}\left(\overline{pqQ(\tau )}\right) \sigma_{xy}\left(\widetilde{pqQ(\tau )}\right) ,
\label{visc}
\eea
where the off-diagonal  stress-energy tensor is
\bea
\sigma_{xy}\left(pqQ(\tau )\right) &=&\sum_{i=1}^N
\frac{p_{i,x}(\tau )\,p_{i,y}(\tau )}{\sqrt{ {\bf p}^2_i(\tau )+m^2_i}}
\cr
&-&
\frac{1}{2}\sum_{i\neq j}^N
q_{ij,x}(\tau )\,\frac{\partial U\left(qQ\right)}{\partial q_{ij,y}}(\tau ),
\label{autosten}
\eea
here ${\bf q}_{ij}={\bf q}_{i}-{\bf q}_{j}$,
and  $U$ is the sum of the color Kelbg potentials defined by Eqs.
(\ref{U-bar}) and (\ref{U-tilde}) with $n=0$.
Examples of the stress-energy-tensor autocorrelation and its antiderivative function are
presented in Fig. \ref{fig:StrVis}. The
shear viscosity is defined by limiting value of the related antiderivative function at $t\rightarrow\infty$.
\begin{figure}[htb]
\hspace*{-5mm}
\includegraphics[width=7.2cm,clip=true]{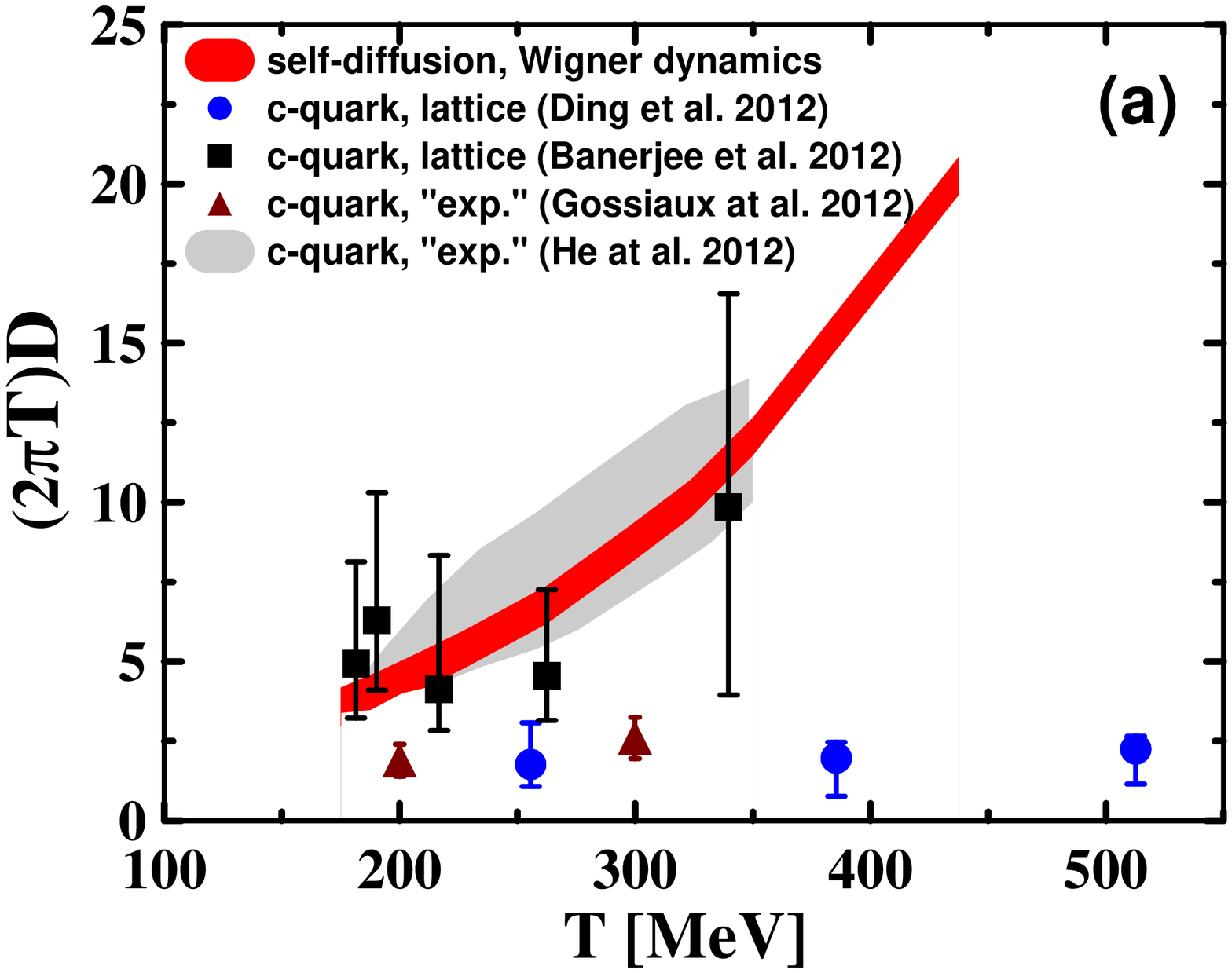}
\\[5mm]
\includegraphics[width=8.cm,clip=true]{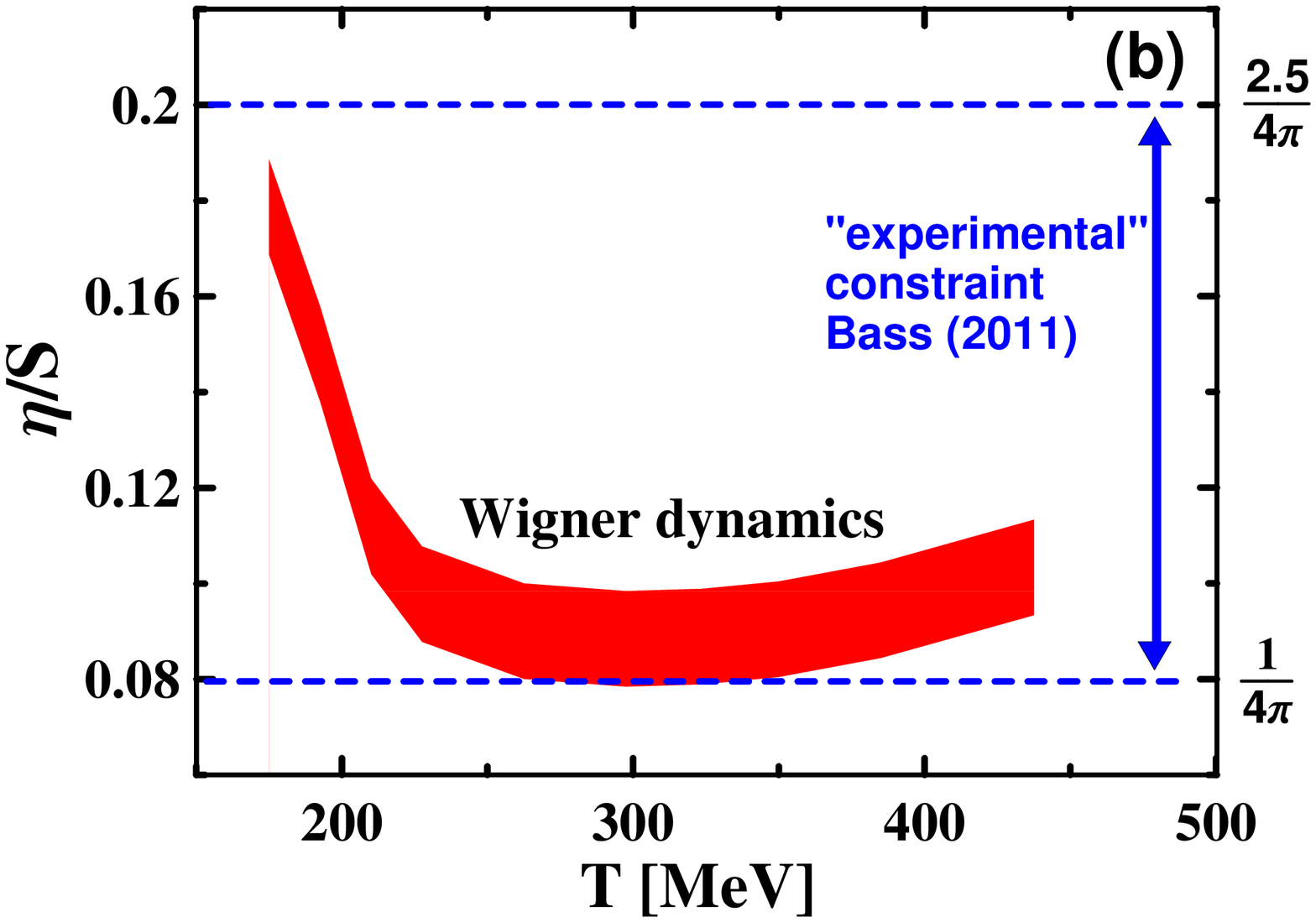}
\caption{(Color online)
Panel a: The self-diffusion constant [self-diffusion, Wigner dynamics]
as a function of temperature,
compared with  heavy-quark diffusion constants predicted by QCD lattice computations
\cite{h-q-diff-latt-11} [c-quark, lattice (Ding et al.2011)] and
\cite{Banerjee:2011ra} [c-quark, lattice (Banerjee et al. 2012)],
and deduced from analysis of experimental data
 \cite{Gossiaux:2012ea} [c-quark, ``exp.'' (Gossiaux at al. 2012)]
and \cite{He:2012df} [c-quark, ``exp.'' (He at al. 2012)].
\\
Panel b: The ratio of shear viscosity to entropy density as a function of temperature
[shaded aria marked ``Wigner dynamics''].
Horizontal dashed lines indicate the range of constraint on the viscosity-to-entropy ratio
deduced from numerous hydrodynamical simulations of heavy-ion experimental data, as
summarized in Ref. \cite{Bass:2011zz}.
}
\label{fig:DifVis}
\end {figure}

The self-diffusion constant and the viscosity-to-entropy ratio
are presented in Fig.~\ref{fig:DifVis} as a function of temperature. The entropy density $s$
is taken from results of our PIMC calculations presented in Fig.~\ref{fig:EOS}.
Our results are presented in the form of bands. The width of the bands represents
the theoretical uncertainty associated with the oscillations  of the
antiderivative functions at large times, see Figs. \ref{fig:VV} and \ref{fig:StrVis}.
Slowly decaying oscillations of the time correlation functions are inherent in liquid-like
systems of  strongly interacting particles in contrast to
exponentially decaying  oscillations in gas-like systems.
As known from hydrogen plasma, these oscillations arise because of quasiclosed chaotic orbits
and are caused by strong interparticle interaction. In liquids these oscillations decay according to a power law
rather than exponentially.
Therefore, extremely long (in time) trajectories are required for more accurate calculations of the
diffusion constant and viscosity. Due to CPU time limitations, we had to stop
our computations before the decay of these oscillations was completed.

Unfortunately, the self-diffusion constant is unavailable from other calculations.
Therefore, we compare it with heavy-quark diffusion constant, anticipating that
these are of the same order of magnitude.
The heavy-quark diffusion constant is available from recent
QCD lattice computations \cite{h-q-diff-latt-11,Banerjee:2011ra},
which are presented in the Panel (a) of Figs.~\ref{fig:DifVis}.
Our results (labeled as ``Wigner dynamics'') well agree with lattice
data of Ref. \cite{Banerjee:2011ra}, while essentially overestimate those
of Ref. \cite{h-q-diff-latt-11}.
The heavy-quark diffusion constant is also available from analysis of
experiments on  heavy-quark quenching in ultrarelativistic heavy ions collisions at RHIC.
For comparison we took two recent works on the subject
\cite{Gossiaux:2012ea,He:2012df}.
Here the results are also rather diverse.
Estimates of Ref. \cite{He:2012df} are well comparable with our result and
lattice data of Ref. \cite{Banerjee:2011ra}, while estimate
of Ref. \cite{Gossiaux:2012ea,He:2012df} is considerably lower and better
conforms to lattice results of Ref. \cite{h-q-diff-latt-11}.


Our results on the shear viscosity are presented in the Panel (b) of Figs.~\ref{fig:DifVis}.
As seen, the obtained values of viscosity are in the range of those
deduced from the analysis of experimental elliptic flow in ultrarelativistic heavy ions collisions at RHIC,
as summarized in Ref. \cite{Bass:2011zz}.
Lattice data on the shear viscosity in the realistic case of the SU(3) group are not available,
an extensive review of theoretical works on viscosity within QCD-motivated models
is done in Ref. \cite{Khvorostukhin:2010aj}. As seen, the minimum of
the viscosity-to-entropy ratio is reached at a temperature above the expected phase
transition rather than at the phase transition point, as is commonly expected.
This minimum turns out to be quite shallow.
The value of the viscosity-to-entropy ratio at the minimum is very
close (from above) to the
lower bound of $\eta/S = 1/4\pi$ for this quantity \cite{Kovtun:2004de}, often referred to as the KSS bound.
With the temperature decrease, i.e. towards the hadronic
phase, the viscosity rapidly rises.

\section{Conclusion}\label{s:discussion}

In this paper we demonstrated that color quantum Monte-Carlo (PIMC) simulations
based on the quasiparticle model of the QGP
are able to reproduce the lattice equation of state at zero baryon chemical potential at
realistic model parameters (i.e. quasiparticle masses and coupling constant)
even near the critical temperature  and also
yields valuable insight into the internal structure of the QGP.
In our simulations we have introduced a new relativistic path integral measure and have developed
a procedure of sampling color quasiparticle variables according to the SU(3) group Haar measure
with appropriate Casimir conditions. Unfortunately, convergence of our calculations
 becomes poor in the range of the expected phase transition
 because the scheme suffers from
jumps between stable and metastable states which  turn out to be
almost equally probable in this range.
Our results indicate that the QGP reveals {\em quantum liquid-like}
(rather than gas-like) properties up to the highest considered temperature of $525 \text{MeV}$.

Short-distance correlations in the computed pair distribution functions
of gluon-gluon and gluon-(anti)quark pairs  displays the formation of clusters.
In case of the gluon-gluon clusters we can even speak about gluon-gluon bound states, i.e. glueballs,
at temperatures just above the phase transition.
The possible existence of medium-modified meson-like bound states was actively
discussed some time ago  \cite{Shuryak03,Brown05}.
Our result supports the conclusion of Ref.
\cite{Koch05} on the absence of $q\overline{q}$ bound states above the temperature of the phase transition.
This finding is in contrast to our previous results on the SU(2)
 group \cite{Filinov:2012pimc,Filinov:2010pimc,Filinov:2012zz,Filinov:2009pimc}.
There well pronounced bound $q\overline{q}$  states were found just above the critical temperature,
which however quickly dissolved with the temperature rise.
This happens because the SU(3) plasma turns out to be essentially denser than the SU(2) one,
which is a consequence of a stronger effective attraction between the constituents.
As a result, possible meson-like bound states in the SU(3) plasma just melt.

The PIMC method is not able to yield  transport properties of the QGP.
A way to access these is to develop a classical color molecular dynamics simulation
 \cite{shuryak1},
where quantum effects are included phenomenologically via a short-range potential.
In contrast to these classical MD simulations \cite{shuryak1}, we have developed a more rigorous approach
 based on the combination of Feynman and Wigner formulations of quantum dynamics.
The basic ideas of this approach have been briefly reported in Ref. \cite{ColWig11}.
In this paper we gave a more detailed description.
In particular, this approach allowed us to calculate the self-diffusion coefficient and the viscosity of the strongly coupled QGP.
Since the self-diffusion constant is unavailable from other calculations,
we compared it with the heavy-quark diffusion constant, anticipating that
these are of the same order of magnitude.
The heavy-quark diffusion constant is available from recent
QCD lattice computations \cite{h-q-diff-latt-11,Banerjee:2011ra}
and also  from an analysis of the heavy-quark quenching in
experiments on ultrarelativistic heavy ions collisions at RHIC.
For comparison we took two recent works on such an analysis
\cite{Gossiaux:2012ea,He:2012df}.
Unfortunately the above mentioned  lattice and heavy-quark-quenching results are rather diverse.
Our self-diffusion constant well agrees with lattice
data of Ref. \cite{Banerjee:2011ra} and  estimates of Ref. \cite{He:2012df},
while essentially overestimates those
of Refs. \cite{h-q-diff-latt-11} and  \cite{h-q-diff-latt-11}.

Our results on the shear viscosity are in the range of those
deduced from the analysis of the experimental elliptic flow in ultrarelativistic heavy ions collisions at RHIC,
as summarized in Ref. \cite{Bass:2011zz}, i.e.  in terms of the viscosity-to-entropy ratio,
$1/4\pi\lsim \eta/S < 2.5/4\pi$, in the temperature range from 170 to 440 MeV.
The minimum of the viscosity-to-entropy ratio is reached at a temperature ($\approx 300$ MeV), above the expected phase
transition rather than at the phase transition point as commonly expected.
This minimum turns out to be very shallow.
The value of the viscosity-to-entropy ratio at the minimum is very
close (from above) to the lower bound of $\eta/S = 1/4\pi$ for this quantity \cite{Kovtun:2004de}, i.e. to the KSS bound.
With the temperature decrease, i.e. towards the hadronic
phase, the viscosity rapidly rises.

Our present analysis is still confined only to the case of zero
baryon chemical potential. Simulations at nonzero baryon chemical potentials are in progress.


We acknowledge stimulating discussions with  P.~Levai, D.~Blaschke, R. Bock,  H.~Stoecker, and D.N.Voskresensky.
Y.B.I. was partially supported by grant of the Russian Ministry of
Science and Education NS-215.2012.2.

\section*{Appendix I: Integration over  SU(3) group Haar measure }

In this appendix we explain details of integration over SU(3) Haar measure $d\mu Q$ in Eq.~(\ref{Gq-def}).
The measure for single color charge in case of the SU(3) group is \cite{LM105,FnInSp,Kell1,Kell2}
\bea
d\mu Q=
d^8Q\,\delta(Q_aQ_a-\breve{q}_2)\,\delta(d_{abc}Q_aQ_bQ_c-\breve{q}_3)
\label{apxdq}
\eea
with summation over $a,b,c=1, \dots , 8$ and constants $d_{abc}$ given in Table 1.
For the SU(N) group the quadratic $\breve{q}_2$ and cubic $\breve{q}_3$ Casimirs are
 $\breve{q}_2=({N}^2-1)C^2$ with $C_2={N}$ for qluons and $C_2=1/2$ for quarks and antiquarks,
 $\breve{q}_3=0$ for gluons and $\breve{q}_3=({N}^2-4)({N}^2-1)/4$ for quarks,  for antiquarks $\breve{q}_3$ has opposite sign. In fact, the normalization constant $c_R$ depends on $\breve{q}_2$ and $\breve{q}_3$ Casimirs.

For random sampling of the $Q$  variable in Monte-Carlo integration in Eq.~(\ref{Gq-def}) we change
to the related canonical Darboux variables for the SU(3) group. The set of the canonical variables
$[\phi\pi]=[(\phi_{\alpha}, \pi_{\alpha}, \alpha=1,2,3]$ is defined by the canonical Poisson bracket
\bea
\left\{A,B\right\}_{PB}=\frac{\partial A}{\partial r}\frac{\partial B}{\partial p}-\frac{\partial A}{\partial p}\frac{\partial B}{\partial r}+
\frac{\partial A}{\partial \phi}\frac{\partial B}{\partial \pi}-\frac{\partial A}{\partial \pi}\frac{\partial B}{\partial \phi}
\label{pois}
\eea
where $r$ and $p$ are conventional coordinate and momentum, respectively,
%
and obey
\bea
\left\{r_\alpha,p_\gamma\right\}_{PB}=\delta_{\alpha\gamma} \qquad    \left\{\phi_\alpha,\pi_\gamma\right\}_{PB}=\delta_{\alpha\gamma}
\label{pb}
\eea
The color variables $Q_a$ form a representation of SU(3). In terms canonical variables their Poison bracket reads
\bea
\left\{Q_a,Q_a\right\}_{PB}=\breve{f}_{abc}Q_c
\label{pbq}
\eea
where $\breve{f}_{abc}$ are the structure constants of SU(3) given in Table 1.

\begin{widetext}
\begin{center}
\begin{table}[htb]\label{tab1}
\caption{
Non-zero $\breve{f}_{abc}$ and $d_{abc}$ constants of the group SU(3)}
\begin{tabular}{|r|c|c|c|c|c|c|c|c|c|}
$\breve{f}_{abc}$ & $\breve{f}_{123}$ & $\breve{f}_{147}$ & $\breve{f}_{156}$ & $\breve{f}_{246}$ & $\breve{f}_{257}$ & $\breve{f}_{345}$ & $\breve{f}_{367}$  & $\breve{f}_{458}$ & $\breve{f}_{678}$ \\
\hline
 values   & $1$ &  $\frac{1}{2}$ &  $-\frac{1}{2}$  &  $\frac{1}{2}$ &  $\frac{1}{2}$ &  $\frac{1}{2}$ &   $-\frac{1}{2}$ &  $\frac{\sqrt{3}}{2}$ &  $\frac{\sqrt{3}}{2}$ \\
\end{tabular}
\begin{tabular}{|r|c|c|c|c|c|c|c|c|c|c|c|c|c|c|c|c|}
$d_{abc}$ & $d_{118}$ & $d_{146}$ & $d_{157}$ & $d_{228}$ & $d_{247}$ & $d_{256}$ & $d_{338}$  & $d_{344}$ & $d_{355}$ & $d_{366}$ & $d_{377}$ & $d_{448}$ & $d_{558}$ & $d_{668}$ & $d_{778}$ & $d_{888}$ \\
\hline
 values   & $\frac{1}{\sqrt{3}}$ &  $\frac{1}{2}$ &  $\frac{1}{2}$  &  $\frac{1}{\sqrt{3}}$ &  $-\frac{1}{2}$ &  $\frac{1}{2}$ &  $\frac{1}{\sqrt{3}}$ &  $\frac{1}{2}$ &  $\frac{1}{2}$ &  $-\frac{1}{2}$ &  $-\frac{1}{2}$ &  $-\frac{1}{2\sqrt 3}$ & $-\frac{1}{2\sqrt 3}$ & $-\frac{1}{2\sqrt 3}$ & $-\frac{1}{2\sqrt 3}$ & $-\frac{1}{\sqrt 3}$ \\
\end{tabular}
\end{table}
\end{center}
\end{widetext}

The explicit transformations to canonical variables are given \cite{LM105,FnInSp,Kell1,Kell2} by expressions:
\bea&&
Q_1=\pi_+\pi_-\cos\phi_1, \qquad \qquad \quad 
\nonumber\\&&
Q_2=\pi_+\pi_-\sin\phi_1, \quad \qquad \qquad 
\nonumber\\&&
Q_3=\pi_1,
\nonumber\\&&
Q_4=C_{++}\pi_+A + C_{+-}\pi_-B, \quad 
\nonumber\\&&
Q_5=S_{++}\pi_+A + S_{+-}\pi_-B,
\nonumber\\&&
Q_6=C_{-+}\pi_-A + C_{--}\pi_+B, \quad 
\nonumber\\&&
Q_7=S_{-+}\pi_-A - S_{--}\pi_+B,
\qquad 
\nonumber\\&&
Q_8=\pi_2
\label{drb1}
\eea
in which we have used definitions:
\bea&&
\pi_+=\sqrt{\pi_3 +\pi_1} \qquad \qquad \qquad \qquad 
\nonumber\\&&
\pi_-=\sqrt{\pi_3 -\pi_1}
\nonumber\\&&
C_{\pm \pm }=\cos\left[\frac{1}{2}\left(\pm \phi_1+\sqrt{3}\phi_2 \pm \phi_3\right)\right] \quad
\nonumber\\&&
S_{\pm \pm }=\sin\left[\frac{1}{2}\left(\pm \phi_1+\sqrt{3}\phi_2 \pm \phi_3\right)\right]
\label{drb2}
\eea
and $A$ and $B$  are given by
\begin{widetext}
\bea&&
A=\frac{1}{2 \pi_3} \sqrt{\left(\frac{J_1-J_2}{3}+ \pi_3+\frac{\pi_2}{\sqrt{3}}\right)
                              \left(\frac{J_1+2J_2}{3}+\pi_3+\frac{\pi_2}{\sqrt{3}}\right)
                              \left(\frac{2J_1+J_2}{3}-\pi_3-\frac{\pi_2}{\sqrt{3}}\right)}
\nonumber\\&&
B=\frac{1}{2 \pi_3} \sqrt{\left(\frac{J_2-J_1}{3} +\pi_3-\frac{\pi_2}{\sqrt{3}}\right)
                              \left(\frac{J_1+2J_2}{3}-\pi_3+\frac{\pi_2}{\sqrt{3}}\right)
                              \left(\frac{2J_1+J_2}{3}+\pi_3-\frac{\pi_2}{\sqrt{3}}\right)}
\label{drb3}
\eea
\end{widetext}
In this expression the set $Q_1, Q_2, Q_3$ forms an SU(2) subgroup with quadratic Casimir
$Q_1^2+Q_2^2+Q_3^2=\pi_3^2$. Let us note that two Casimirs depend only on
$J_1$ and  $J_2$. They can be computed using the values given in Table 1 as
\bea
Q_a Q_a&=&\frac{1}{3}\left(J_1^2+J_1 J_2+J_2^2\right) \qquad
\nonumber\\
d_{abc}Q_a Q_b Q_c&=&\frac{1}{18}\left(J_1-J_2\right)\left(J_1+2J_2\right)\left(2J_1+J_2\right)
\hspace*{6mm}
\label{drb4}
\eea

The phase space color measure for SU(3), given in Eq. (\ref{apxdq}) can be transformed to the new coordinates
through use of Eq. (\ref{drb1}) and evaluation of the Jacobian
\bea
\left|\frac{\partial Q}{\partial(\phi,\pi)}\right|=  \frac{\sqrt 3}{48}J_1J_2(J_1+J_2)
\label{jac}
\eea
Then the measure reads
\bea
d\mu Q &=&  d \phi_1 d \phi_2 d \phi_3 d \pi_1 d \pi_2 d \pi_3 d J_1 d J_2
\frac{\sqrt 3}{48}J_1J_2(J_1+J_2)
\nonumber\\&\times&
\delta \left(\frac{1}{3}((J_1)^2+J_1 J_2+(J_2)^2)-\breve{q}_2\right)
\nonumber\\&\times&
\delta \left(\frac{1}{18}(J_1-J_2)(J_1+2J_2)(2J_1+J_2)-\breve{q}_3\right)
\label{meas}
\eea
Since the two Casimirs are independent, the $\delta$-functions fix both $J^1$ and $J^2$.
After integration over $J^1$ and $J^2$  the Eq.~(\ref{meas}) gives a proper canonical volume element
$d \phi d \pi$.  Thus applying Metropolis algorithm to $\phi \pi$ variables we can construct Markovian chain
in $\phi \pi$ phase space and obtain random  color variables $Q$ for calculation partition function according to the SU(3) group Haar measure with two Casimir  conditions.

\section*{Appendix II: Lebesque-Dirac delta theorem}

Let $f$ be a summable function of  real argument such that $\int f(x)dx =I_0$. \\
Then $Mf(M(x-x')) \rightarrow I_0
\delta(x-x')$ when $M\rightarrow \infty$.

Proof: Let $\varphi $ be any test function.
Then
\begin{eqnarray*}
&&\lim_{M\rightarrow \infty} \int M f\left(M(x-x') \right) \varphi(x)  dx 
\cr
&=&
\lim_{M\rightarrow \infty} \int f(s)\varphi(s/M+x') ds
\cr
&=&\int  f(s) (\lim_{M\rightarrow \infty} \varphi(s/M+x')) ds 
\cr
&=&
\varphi(x')\int f(s) ds = I_0\varphi(x')
\end{eqnarray*}


\bibliographystyle {apsrev}

\end{document}